\documentclass[referee,pdflatex,sn-basic]{sn-jnl}

\jyear{2021}%

\theoremstyle{thmstyleone}%
\theoremstyle{thmstyletwo}%
\usepackage{cancel}
\theoremstyle{thmstylethree}%
\usepackage{rotating}
\usepackage{graphicx}
\usepackage{capt-of}
\usepackage{booktabs}
\usepackage{varwidth}
\usepackage{subcaption}

\DeclareUnicodeCharacter{2212}{-}

\begin{document}

\title[Substitutional Alloying Using Crystal Graph Neural Networks]{Substitutional Alloying Using Crystal Graph Neural Networks}


\author[1]{\fnm{Dario} \sur{Massa}}

\author[1]{\fnm{Daniel} \sur{Cieśliński}}

\author[1]{\fnm{Amirhossein} \sur{Naghdi}}

\author*[1]{\fnm{Stefanos} \sur{Papanikolaou}}\email{stefanos.papanikolaou@ncbj.gov.pl}

\affil*[1]{\orgdiv{NOMATEN Centre of Excellence}, \orgname{National Centre for Nuclear Research}, \orgaddress{ uł. Andreja Sołtana 7, \city{Otwock}, \country{Poland}}}


\abstract{

Materials discovery, especially for applications that require extreme operating conditions, requires extensive testing that naturally limits the ability to inquire the wealth of possible compositions.  Machine Learning (ML) has nowadays a well established role in facilitating this effort in systematic ways. The increasing amount of available accurate DFT data represents a solid basis upon which new ML models can be trained and tested. While conventional models rely on static descriptors, generally suitable for a limited class of systems, the flexibility of Graph Neural Networks (GNNs) allows for direct learning representations on graphs, such as the ones formed by crystals. We utilize crystal graph neural networks (CGNN) to predict crystal properties with DFT level accuracy, through graphs with encoding of the atomic (node/vertex), bond (edge), and global state attributes. In this work, we aim at testing the ability of the  CGNN MegNet framework in predicting a number of properties of systems previously unseen from the model, obtained by adding a substitutional defect in bulk crystals that are included in the training set. We perform DFT validation to assess the accuracy in the prediction of formation energies and structural features (such as elastic moduli). Using CGNNs, one may identify promising paths in alloy discovery. 

}




\maketitle





\section{Introduction}\label{sec1}

The use of machine learning (ML) \cite{ML1,ML2} methods in material science to accelerate materials discovery\cite{matdisc} is at the base of the so-called material informatics (MI) \cite{MI1,MI2,MI3,MI4,MI5}. By training ML models on large databases, such as OQMD or the Materials Project high-throughput electronic structure calculation databases~\cite{DB1,DB2,DB3,DB4,DB5}, the goal is to achieve predictions of material properties with quantum accuracy.

As in statistical mechanics with the need for identifying appropriate order parameters of novel phases and structures, the key challenge in ML algorithms is to identify effective system descriptors that can function as structure identifiers. 
A large variety of descriptors have been proposed, including fixed-length feature vectors of material elemental or electronic properties \cite{fixlen_feat_1,fixlen_feat_2,fixlen_feat_3}, as well as structural descriptors, based on rotational and traslational invariant transformations of atomic coordinates, like the Coulomb matrix \cite{CoulombMat}, atom-centered symmetry functions (ACSFs) \cite{ACSFs}, social permutation invariant coordintes (SPRINT) \cite{SPRINT}, smooth overlap of atomic positions (SOAP) \cite{SOAP} and global minimum of root mean-square distance \cite{globalmin}.
However, these solutions are often system-specific, and are not suitable for vast compositional and structural space exploration.

For this reason, a topic of fervent interest in the materials science community is the use of graph neural networks (GNNs) \cite{GNN_1,GNN_2}, {which allow} to learn representations directly and in a flexible way, focused on molecular systems \cite{GNNmol_1,GNNmol_2,GNNmol_3,GNNmol_4,GNNmol_5,GNNmol_6}, surfaces \cite{GNNsurf1,GNNsurf2,GNNsurf3} and periodic crystals\cite{GNNmol_2,GNNcry1,GNNcry2,GNNcry3,GNNcry4,GNNcry5,GNNcry6,GNNcry7,GNNcry8}. {GNNs can be regarded as the generalization of convolutional neural networks (CNN) to graph-structured data, from which the internal materials representations can be learned and used for prediction of target properties \cite{NatureCommun}; even though larger amounts of data is required with respect to conventional ML models, GNNs take advantage of the unambiguous physics-guided real-space local associations between the system's degrees of freedom, hence for any type of atomic crystalline structure  \cite{Elsevier}.}
The common idea of GNN-based models is to represent atoms as nodes {(V)} and their {chemical bonds} as edges {(E)} in a graph {G(V,E)}, which can be fed to a trained neural network to create node-level embeddings  {(learned representations of each atom in its individual chemical environment)} through convolutions with neighbouring nodes and edges \cite{benchmarkingpaper}. {Therefore, given a set of learnable weights (W), and (y) a target material property, the GNN model reformulates the prediction task as the mapping $f(G:W)\rightarrow$\,y.}

{A direct benefit of the crystal material GNN-converted graph encoding is the naturally derived vector characterization of the atoms and edges \cite{RSC}. The work of Xie \textit{et al.} \cite{GNNcry1} represents the pioneering example of a crystal graph convolutional neural network (CGCNN) architecture, which has been later extended in the iCGCNN by Park \textit{et al.} \cite{Park} to include 3-body correlations on neighbouring atoms, information on the Voronoi tasselated structure and an optimized chemical representation of interatomic bonds in the crystal graphs. }

For the discovery of new materials, one may take various exploring paths, involving high-throughput computational \cite{matdisc} and experimental \cite{matdisc_exp} methods. However, the combined approach of machine-learning methods and compositional manipulation, has very quickly acquired a well established role in materials science, and it is applied in a wide range of property optimization searches like for zinc blende semiconductors \cite{zinc}, perovskites \cite{nature-perovsk, perov1, perov2, perov3, perov4,perov5} and others \cite{compos_manip1,compos_manip3,compos_manip4,MLdef1,MLdef2,MLdef3,MLdef4,MLdef5}.

In this work, we utilize a particular improvement of the originally proposed \cite{GNNcry1} CGCNN model, the MatErials Graph Network (MEGNet) model from Chen, Ye and coworkers \cite{GNNcry8}, introduced in Sec.(\ref{subsubsec:MEGNet}), that has the merit of being developed and tested both on molecules and crystals, with the possibility of defining global state attributes {including temperature, pressure and entropy}. We test the capabilities of graph networks to predict the properties of single-atom substitutionally defected crystals with the MEGNet model. After considering a pre-trained model on the Materials Project (MP) database (Sec.(\ref{subsubsec:pretrained})), we focus on the formation energies, bulk and shear moduli predictions, both comparing the results obtained in datasets of similarly defected structures (Sec.(\ref{subsec:Adefectinmatrices})) and the effects of almost all the possible single-atom defects in the same matrices (Sec.(\ref{subsec:Defectsinamatrix})). To validate the predictions, as described in Sec.(\ref{subsec:validation}) and Sec.(\ref{subsec:val_results}), we perform Density Functional Theory (DFT) calculations, and we find CGNNs have both a great potential, but also limitations in predicting properties of defected bulk crystals, and promote materials discovery. 

\section{Methods}\label{sec:methods}

\subsection{Machine Learning framework} \label{subsec:MLframework}
\subsubsection{MEGNet description} \label{subsubsec:MEGNet}
In the present work, we utilize the MEGNet model\cite{GNNcry8}. The reasons for this choice lie in the structure and performance of the model: 
\begin{enumerate}
    \item It is characterized by a low number of attributes, one for the atom (atomic number) and one for the bond (spatial distance), but MEGNet outperforms previous graph-based models\cite{GNNcry8}, as CGCNN\cite{GNNcry1} and MPNN\cite{GNNmol_1}, with higher number of attributes, as well as SchNet\cite{GNNmol_2}, with a similar low number.
    \item the MEGNet framework includes a global state attribute, essential for state-property relationship predictions in materials,
    \item The graph network construction of MEGNet has been developed and tested for both molecules and crystals
\end{enumerate} 
Here, we limit ourselves to present the main features of the model, but for a more exhaustive explanation we recommend the reader to the original work of Chen~\textit{et al.}\cite{GNNcry8} and references there-in. In particular, given a graph $G(E,V,\textbf{u})$, where

\begin{table}[tbh]
\centering
\begin{tabular}{ | l | r | r |}
    \hline
    Parameter & Value & Short Description \\ \hline \hline
    \texttt{nfeat\_node}  & 94 & number of atom features\\ \hline
    \texttt{nfeat\_global}  & 2 & number of state features \\ \hline
    \texttt{ngauss\_centers}  & 110 & number of gaussians\\ \hline
    \texttt{converter\_cutoff} & 4 & cutoff radius \\ \hline
    \texttt{megnet\_blocks} & 3& number of MEGNetLayer blocks\\ \hline
    \texttt{optimizer} & Adam & optimizer of the model weights\\ \hline
    \texttt{lr} & 1e-3& learning rate\\ \hline
    \texttt{n1} & 64& number of hidden units in layer 1\\ \hline
    \texttt{n2} & 32& number of hidden units in layer 2\\ \hline
    \texttt{n3} & 16& number of hidden units in layer 3\\ \hline
   \end{tabular}
    \caption{Parameters from the pre-trained MEGNet model.}
    \label{table:GNNparams}
\end{table}

\begin{itemize}
    \item $V$ is the set of $N^v$ atomic attribute vectors $\textbf{v}_i$;
    \item $E= \Big\{ \big( \textbf{e}_k, \textbf{r}_k, \textbf{s}_k \big) \Big\}_{k=1..N}^e$ is the set of $N^e$ bond attribute vectors, with $\textbf{r}_k$ and $\textbf{s}$ being the indexes of the connected atoms;
    \item $\textbf{u}$ is the global state attribute vector; 
\end{itemize}
the role of graph network is to recursively update a input graph $G(E,V,\textbf{u})$ to an output graph $G(E', V', \textbf{u'})$, with a progressive and inclusive information flow going from bonds to atoms, and finally to the global state. In particular, first the attributes of each bond are updated through a function $\phi_e$, applied on the concatenation of the self-attributes, the ones of the connecting $\textbf{v}_{s_k}$ and $\textbf{v}_{r_k}$ atoms, and of the global state $\textbf{u}$, as in
\begin{equation}
\textbf{e}_k'=\phi_e \big( \textbf{v}_{s_k}\bigoplus \textbf{v}_{r_k}\bigoplus \textbf{e}_{k}\bigoplus  \textbf{u}  \big)
\end{equation}
The update of atomic attributes involves the average over i-th atom connecting bonds $\Bar{\textbf{v}}_i^e=\dfrac{1}{N^e}\sum_{k=1}^{N_i^e} \big\{  \textbf{e}'_k  \big\}_{r_k=i}$, the i-th atom self-attributes $\textbf{v}_i$ and the global state ones $\textbf{u}$, as in \begin{equation}
\textbf{v}'_i=\phi_v \big( \Bar{\textbf{v}}_i^e \bigoplus \textbf{v}_i\bigoplus \textbf{u}\big)
\end{equation}
Finally, an information flow from all three attribute groups is involved in the update of the global state attibutes, as in \begin{equation}
    \textbf{u}'=\phi_u \big( \Bar{\textbf{u}}^e \bigoplus \Bar{\textbf{u}}^v \bigoplus \textbf{u} \big)
\end{equation}
where $\Bar{\textbf{u}}^e=\dfrac{1}{N^e}\sum_{k=1}^{N^e} \big\{  \textbf{e}'_k  \big\}$ and $\Bar{\textbf{u}}^v=\dfrac{1}{N^v}\sum_{i=1}^{N^v} \big\{  \textbf{v}'_i  \big\}$. \vskip 0.2 pt As mentioned before, for our systems of interest, namely periodic crystals, the atomic number is the only node attribute. For bonds, the spatial distance is expanded in a Gaussian basis set, centered at a linearly spaced $r_0$ locations between $r_0=0$ and $r_0=r_{\text{cut}}$, and characterized by a given width $\sigma$\footnote{Therefore of shape $\text{exp}\big( -(r-r_0)^2 / \sigma^2 \big)$.}. Finally, the global state is simply a two zeros placeholder for global information exchange. 

\begin{figure}[bth]
    \centering
    \includegraphics[width=\textwidth]{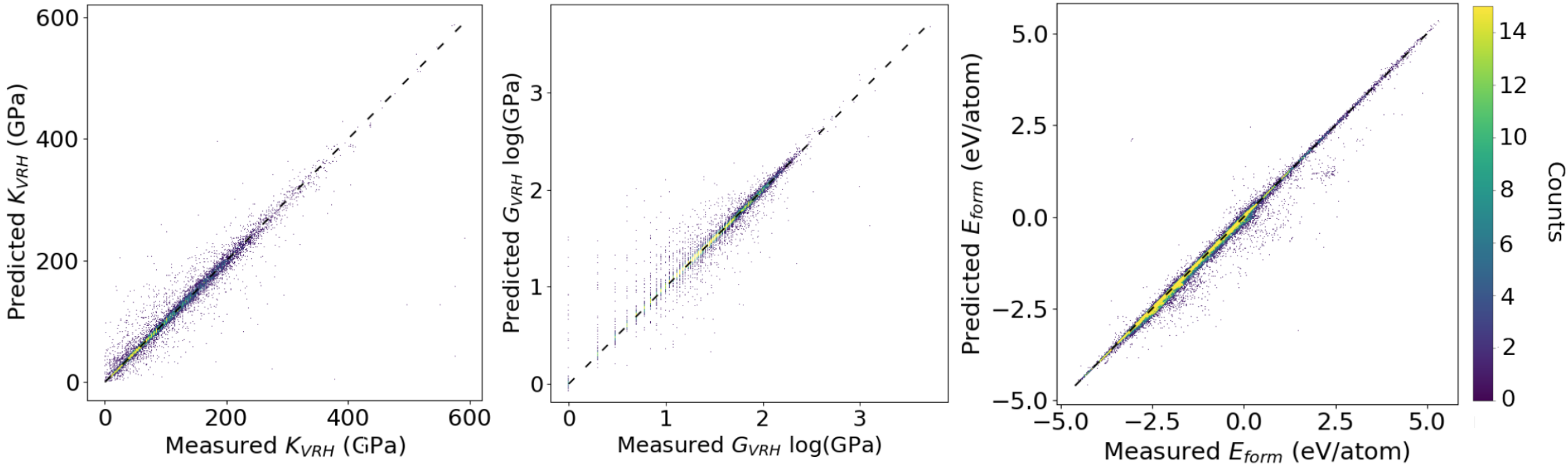}
    \caption{Parity plots for the pre-trained model on the MP dataset. The plots involve the predictions on the bulk modulus ($\text{K}_{\text{VRH}}$), the shear modulus ($\text{G}_{\text{VRH}}$) and the formation energy ($\text{E}_{\text{form}}$). }
    \label{fig:parity}
\end{figure}

\subsubsection{Data collection} \label{subsubsec:data}
We consider crystal structures collected through the Python Materials Genomics interface (pymatgen) \cite{pymatgen} to the Materials Application Programming Interface from Materials Project \cite{DB2}. When creating the dataset, there were 126301 structures in the database, that had formation energy ($\text{E}_{\text{form}}$) property, and 13102 structures that had bulk modulus ($\text{K}_{\text{VRH}}$) and shear modulus ($\text{G}_{\text{VRH}}$) properties.

\subsubsection{Pre-trained model} \label{subsubsec:pretrained}
Our focus in this work is on the prediction capabilities of  MEGNet for \emph{minimally} defected systems, that are clearly not in the training database, given the training of a dataset of undefected structures. In order to do so, we consider the substitution of a single atom in a supercell, hoping that CGNN training captures atomic similarities, based on combinations of atomic radii, valence electrons, and other atomic properties. Table(\ref{table:GNNparams}) shows some of the paramaters of the model, and a more complete list can be found at the default implementation of the class~\cite{megnetmodel}

\begin{table}[tbh]
\centering
\begin{tabular}{ | l | r | r |}
    \hline
    Property & MAE \\ \hline \hline
    $\text{K}_{\text{VRH}}$ & 6.143 GPa \\ \hline
    $\text{G}_{\text{VRH}}$ & 10.489 GPa\\ \hline
    $\text{E}_{\text{f}}$ & 0.029 eV/atom\\ \hline
   \end{tabular}
    \caption{MAEs of the model for the prediction of the bulk modulus ($\text{K}_{\text{VRH}}$), shear modulus ($\text{G}_{\text{VRH}}$) and formation energy ($\text{E}_{\text{form}}$). }
    \label{table:tableparity}
\end{table}

 We report parity plots of Fig.(\ref{fig:parity}) for all three properties of interest in this study: bulk modulus ($\text{K}_{\text{VRH}}$), shear modulus ($\text{G}_{\text{VRH}}$) and formation energy ($\text{E}_{\text{form}}$). To evaluate the model accuracy in predicting the properties of interest for the present study, the mean-absolute error (MAE) is used as the evaluation metric. Table(\ref{table:tableparity}) presents the MAE values for each predicted property over the dataset, which provides insights into the pre-trained model performance.

\subsection{Validation with DFT}\label{subsec:validation}
We verify the accuracy of the model's predictions for system properties such as bulk modulus $\text{K}_{\text{VRH}}$, shear modulus $\text{G}_{\text{VRH}}$ and formation energy $\text{E}_{\text{form}}$, after single-atom substitution is implemented in $2\times2\times2$ supercells. We perform DFT calculations with \texttt{Quantum Espresso} (QE) \cite{QE1,QE2,QE3} and its \texttt{THERMO\_PW} \cite{THMPW} driver for the calculation of structural properties. Pseudo-Potentials for all involved atomic species are Ultra-Soft and with Perdew-Burke-Ernzerhof (PBE) \cite{PBE} functional. The Methfessel-Paxton smearing~\cite{Smearing}  has been introduced to correctly investigate metallic systems, and the calculations have been set as spin-polarized, for possible non-zero magnetization effects. Convergence is checked on the number of k-points,  plane-wave cutoff energy, and also, the energy smearing spreading (\texttt{degauss} parameter in QE), for each case, after a preliminary variable-cell relaxation of  pure crystals, and then,  further fixed-cell relaxation with optimal parameters, for final equilibrium bulk structures. The common acceptance threshold in the variation of the total energy upon parameter change is set at $10^{-5}$ Ry.  Forces and total energy convergence thresholds for ionic minimization are set to a common value of $10^{-5}\text{a.u.}$ and $10^{-6}\text{a.u.}$ respectively. After optimization of pure crystals, fixed-cell relaxation is performed on supercells with single-atom substitutions, and its structural properties are then extracted through the \texttt{THERMO\_PW} driver. \vskip 0.2 pt The computation of formation energies for validation purposes is performed on the case of single-atom substitutional defects (D) applied on pure bulk crystals (M), as follows, \begin{equation}
    \text{E}_{\text{form}} \big( \text{M}_{1-\text{x}} \text{D}_{\text{x}} \big) = \text{E}\big( \text{M}_{1-\text{x}} \text{D}_{\text{x}} \big) - \big( 1-\text{x}\big) \text{E}\big(\text{M}\big) - \text{x}\text{E}\big( \text{D} \big)
\end{equation} 
where \text{x} is the atomic fraction of  substitutional defects, $\text{E}\big( \text{M}_{1-\text{x}} \text{D}_{\text{x}} \big)$ is the total energy per atom of the compound, while $\text{E}\big(\text{M}\big)$ and \text{E}\big(\text{D}\big) are the total energies per atom of the precursor species that compose it; the latter energies are obtained by relaxing the ground-state lattices of these species using the same aforementioned QE parameters as for the compound, as necessary to ensure computational consistency in the evaluation of accurate values for the defects formation energies.


\section{Results: Property predictions for substitutional alloying with CGNN}\label{sec:Results}
There is a large variety of ways to systematically evaluate the effect of substitutional alloying on the properties of crystals \cite{KaxirasBook}. Here, we focus on two key questions:
\begin{enumerate}
    \item Data Science of Defects: What is the effect of substituting the same defect in a large variety of systems? 
    \item Is there qualitative and quantitative effects from atom-substituting various elemental defects in the same host crystalline matrix? 
\end{enumerate}
While it has not yet been possible to perform an exhaustive search of the kind, in this work, for the first part, the basin of host systems is represented by the Materials Project crystals dataset introduced in the previous sections \cite{DB2}, while for the latter part, the host systems are pure metallic bulk crystalline supercells of Al, Ni, Mo and Au. 

\subsection{An elemental defect seeing a wealth of different crystalline environments} 
\label{subsec:Adefectinmatrices}

First, we focus on the prediction of  system properties, where the systematic substitutional defective process is applied using the same replacement atom, on randomly selected sites of crystals that exist in the the Materials Project crystals dataset \cite{DB2}. The aim is to explore how material properties change after single-atom substitution, evaluating deviations from original pure crystalline predictions, and how they depend on the specific substitution. For we consider three elemental cases: Rb, Mn and H, as the key replacement atoms that we will mutually compare. The reason for the choice lies in the drastic elemental differences among their unique characteristics and the assumption that, on the base of these, we might be able to gain deeper understanding on how the model behaves in the predictions of previously unseen defect-induced changes in the system properties, basing on its learned notion of local environment. Table(\ref{tab:elemental_differences}) reports, as an example, the values of the atomic weight, radius and electronic configuration of each considered replacement atom.

\begin{table}[]
    \centering
    \begin{tabular}{|l|r|r|r|} \hline 
          &Weight(u) &Radius (pm) &El. Configuration \\ \hline \hline
         H& 1.008 & 53 & 1$\text{s}^1$  \\ \hline
         Mn& 54.938 & 161 & [Ar] 4$\text{s}^2$ 3$\text{d}^5$ \\ \hline
         Rb& 85.468 & 265 & [Kr] 5$\text{s}^1$ \\\hline
    \end{tabular}
    \caption{Differences in the atomic properties of the three elements considered for the single-atom substitutional process. As an example, we report here the atomic weight, radius and electronic configuration.}
    \label{tab:elemental_differences}
\end{table}

\begin{figure}[H]
\centering
\includegraphics[width=92mm]{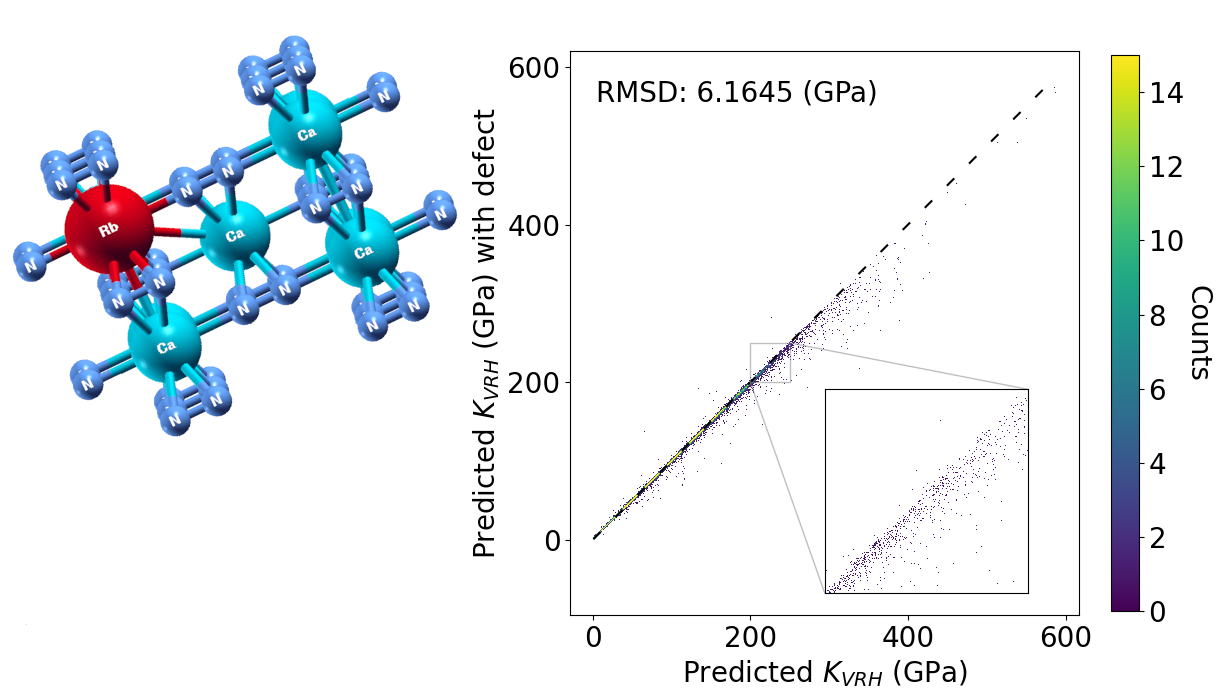}
\captionof{figure}{Predicted bulk moduli in the Rb-defected MP crystals with respect to their prediction in non-defected ones. An example structure from the MP dataset is shown, CaN2, in which one atom of Ca has been replaced with a Rb atom. Here only the case of Rb-defected systems is shown, due to its largest RMSD among the set of considered defects. Similar plots for the Mn- and H-defected systems are contained in the Supplementary Materials.} . 
\label{fig:BMpred-1case}
\end{figure}

In Fig.(\ref{fig:BMpred-1case}), we display the predictions of MEGNet for the bulk moduli of Rb-defected systems with respect to the original, pure ones. As contained in Table(\ref{table:BMpred-1case}), this kind of substitutional defect causes the largest root-mean-squared (RMSD) deviations among the three considered atomic species, with a value of approximately $6.2$ GPa. In this section, we only report plots referring to the largest RMSD cases, but analogous plots can be found in the Appendix Sec.(\ref{secA1}).

\begin{table}[htb]
\centering
\begin{tabular}{ | l | r | r |}
    \hline
    Defect & RMSD (GPa) \\ \hline \hline
    Rb & 6.1645 \\ \hline
    Mn & 3.1274 \\ \hline
    H & 1.2975 \\ \hline
   \end{tabular}
    \caption{RMSD (GPa) for the prediction of the bulk modulus ($\text{K}_{\text{VRH}}$) in Rb, Mn, and H single-atom substitutionally defected systems with respect to the non-defected ones.}
    \label{table:BMpred-1case}
\end{table}
Also the shear moduli predictions show the largest RMSD for the case of a Rb-defect, with a value of $0.0628$ log(GPa)\footnote{As it can be visible from an analogous plot in the Supplementary Material, without a log-log scale the deviations cannot be correctly estimated. This is the reason why we decided to report the $\text{G}_{\text{VRH}}$ prediction plots and RMSD values in these units.}. The results are shown in Fig.(\ref{fig:SMpred-1case}) and on  Table(\ref{table:SMpred-1case}). According to the model, it seems that both $\text{K}_{\text{VRH}}$ and $\text{G}_{\text{VRH}}$ upon substitution of a Rb atom, are showing a tendency to decrease their value, respectively, implying an increase in their compressibility and decrease in hardness. It is worth noticing that even though the only physical atomic feature that the model exploits is the atomic number, the prediction of larger changes in the structural properties for involved defects with larger radii can be regarded as a reasonable one. 

\begin{table}[htb]
\begin{minipage}[b]{0.40\linewidth}
\centering
\begin{tabular}{ | l | r | r |}
    \hline
    Defect & RMSD (log(GPa)) \\ \hline \hline
    Rb & 0.0628 \\ \hline
    Mn & 0.0350 \\ \hline
    H & 0.0435 \\ \hline
   \end{tabular}
    \caption{RMSD (log(GPa)) for the prediction of the shear modulus ($\text{G}_{\text{VRH}}$) in Rb, Mn, and H single-atom substitutionally defected systems with respect to the non-defected ones.}
    \label{table:SMpred-1case}
\end{minipage}\hfill
\begin{minipage}[b]{0.56\linewidth}
\centering
\includegraphics[width=57mm]{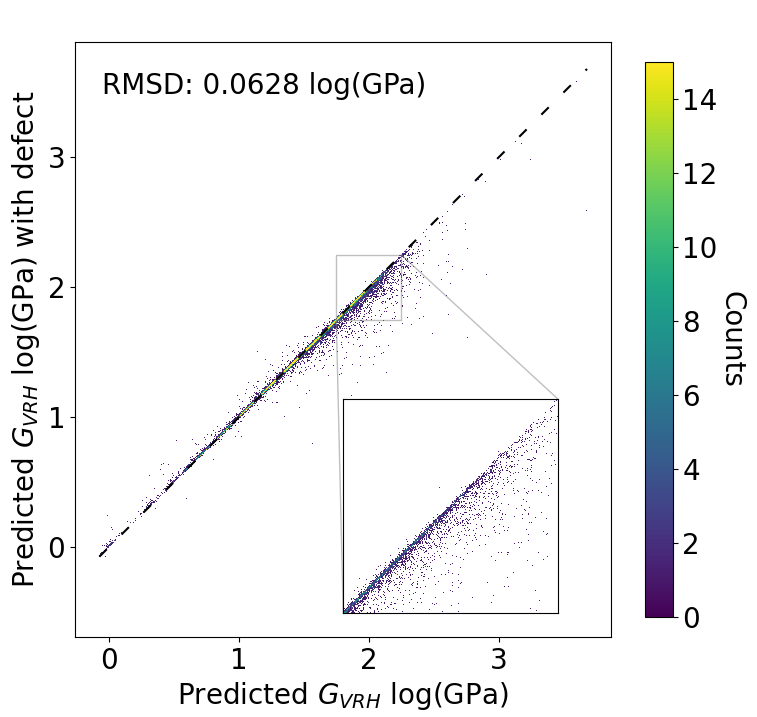}
\captionof{figure}{Predicted shear modulus in the Rb-defected MP crystals with respect to the prediction in non-defected ones. Here only the case of Rb-defected systems is shown, due to its largest RMSD among the set of considered defects. Similar plots for the Mn and H-defected systems are contained in the Supplementary Materials.}
\label{fig:SMpred-1case}
\end{minipage}
\end{table}

Furthermore, in Fig.(\ref{fig:FEpred}), we show the predicted deviations in the formation energy of the MP-defected samples for the case of elemental H, and also for all three elements in Tab.(\ref{table:FEpred-1case}). As it can be seen, the deviations are all in the order of ~$0.02$ eV/atom, and are progressively less prominent going from H, to Rb  and Mn. 
\begin{table}[htb]
\begin{minipage}[b]{0.40\linewidth}
\centering
\begin{tabular}{ | l | r | r |}
    \hline
    Defect & RMSD (eV/atom) \\ \hline \hline
    Rb & 0.0189 \\ \hline
    Mn & 0.0150 \\ \hline
    H &  0.0231\\ \hline
   \end{tabular}
    \caption{RMSD (eV/atom) for the prediction of the formation energy ($\text{E}_{\text{form}}$) in Rb, Mn, and H single-atom substitutionally defected systems with respect to the non-defected ones.}
    \label{table:FEpred-1case}
\end{minipage}\hfill
\begin{minipage}[b]{0.56\linewidth}
\centering
\includegraphics[width=57mm]{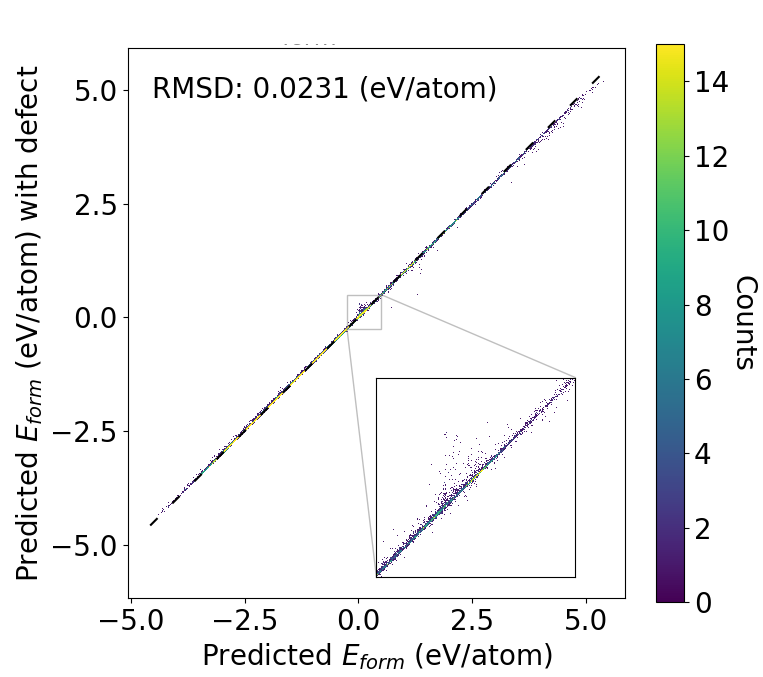}
\captionof{figure}{Predicted formation energy in the H-defected MP crystals with respect to the prediction in non-defected ones. Here only the case of H-defected systems is shown, due to its largest RMSD among the set of considered defects. Similar plots for the Rb and Mn-defected systems are contained in the Supplementary Materials.}
\label{fig:FEpred}
\end{minipage}
\end{table}

\subsection{A crystalline matrix seeing a wealth of different elemental defects}\label{subsec:Defectsinamatrix}
\begin{figure}[H]
    \centering
    \includegraphics[width=\textwidth]{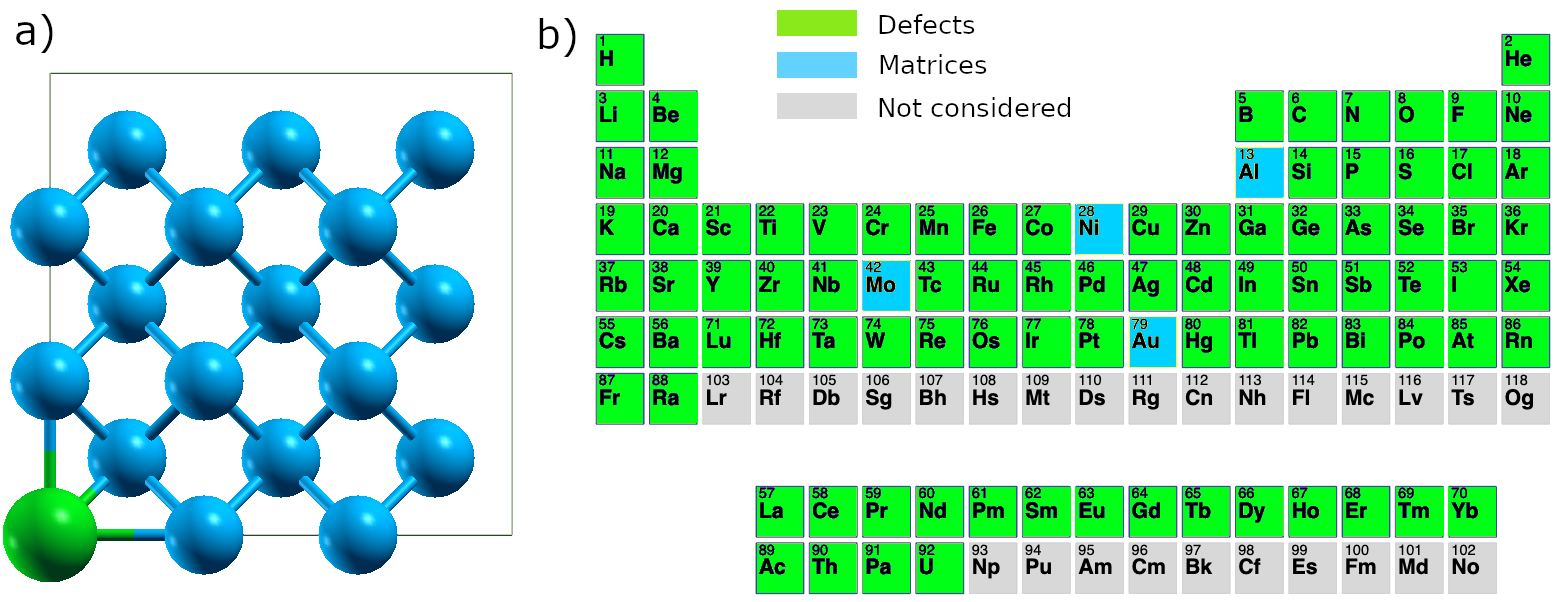}
    \caption{Periodic table plot (b) explaining the process of selecting a set of host matrices (in light blue) and substitutional defects (in green). To a given a selected host matrix, the non-selected ones represent defects too. The host matrices are $3\times3\times3$ supercells of the highlighted species (a).}
    \label{fig:firstpt}
\end{figure}

The case of systematically changing the single-atom substitutional defect species in the same crystalline matrix is complementary to the prior addressed one. We focus on pure metallic bulk host matrices, namely Al, Ni, Mo and Au, even though the test could be done on any crystalline matrix. In Fig.(\ref{fig:firstpt}), we show the key aspects of the performed calculation, with the supercell shown on the left (a), and  the elements considered on the right (b). The compositional space considered for the single-atom substitution covers almost the entire periodic table, and also comprises the species highlighted as host matrices when they are not selected as such. Even though our focus is on simple systems, the process can be insightful on the capabilities of the model to learn and predict, with minimal input, physio-chemical  trends throughout the periodic table. 

\begin{figure}[H]
    \centering
    \includegraphics[width=\textwidth]{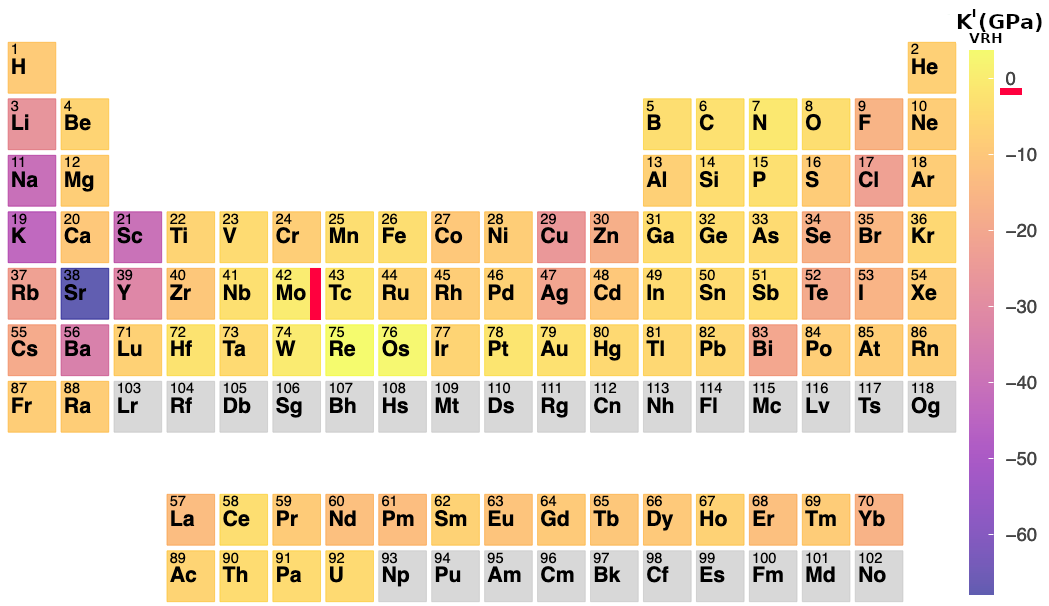}
    \caption{Predicted bulk modulus variation ($\text{K'}_{\text{VRH}}$) for a single-atom substitutionally defected Mo supercell with respect to the undefected one, for each possible defect atomic specie from the provided periodic table. Similar plots for Al, Au and Ni supercells are provided in the Supplementary Materials. The red flag highlights the zero relative difference, meaning the pure Mo matrix selection. }
    \label{fig:BMpred_ptable}
\end{figure}

We show periodic table plots for the properties of interest, $\text{K}_{\text{VRH}}$, $\text{G}_{\text{VRH}}$ and $\text{E}_{\text{form}}$, for the host matrix which displays the largest deviations in the properties, this is the case of Mo, but analogous plots for the other matrices are included in the Supplementary Material (SM). With the help of the visualization style, we characterize the predictions of the effect on the chemical distance between the substitutional defects and the host matrix on the properties of interest, and how it correlates with the well known trends along the periodic table. In Fig.(\ref{fig:BMpred_ptable}), we show the prediction of the host supercell bulk modulus variation when defected with one of the elements from the periodic table. In particular, in the plot we refer to 
\begin{equation}    \text{K'}_{\text{VRH}}=\text{K}^{\text{Mo(X)}}_{\text{VRH}}-\text{K}^{\text{Mo}}_{\text{VRH}}
\end{equation}
where $\text{K}^{\text{Mo(X)}}_{\text{VRH}}$ is the bulk modulus of the Mo matrix defected by atom X, and $\text{K}^{\text{Mo}}_{\text{VRH}}$ its value for pure Mo (labelled with a red flag in the figure). 

Even though Sr represents an outstanding outlier in decreasing the bulk modulus of the Mo crystal, at this scale it is still possible to appreciate how the variation happens along the periods: for the 3d, 4d and 5d transition metals from 3rd to 12th group, the defect-induced variations are mainly small, while a tendency to increase is observed, in modulus, in the post-transition metals and, remarkably, in the alkali and alkaline-earth. This behaviour can be interpreted in terms of the well known variations of the bulk modulus along the periodic table of elements, which seem correlated with respect to the defect-induced ones contained in this plot. The effect of substitutional alloying species which, in their bulk and standard temperature and pressure (STP) conditions, show a lower bulk modulus, is, as a tendency, to lower the bulk modulus of their host system. 

\begin{figure}[H]
    \centering
    \includegraphics[width=0.65\textwidth]{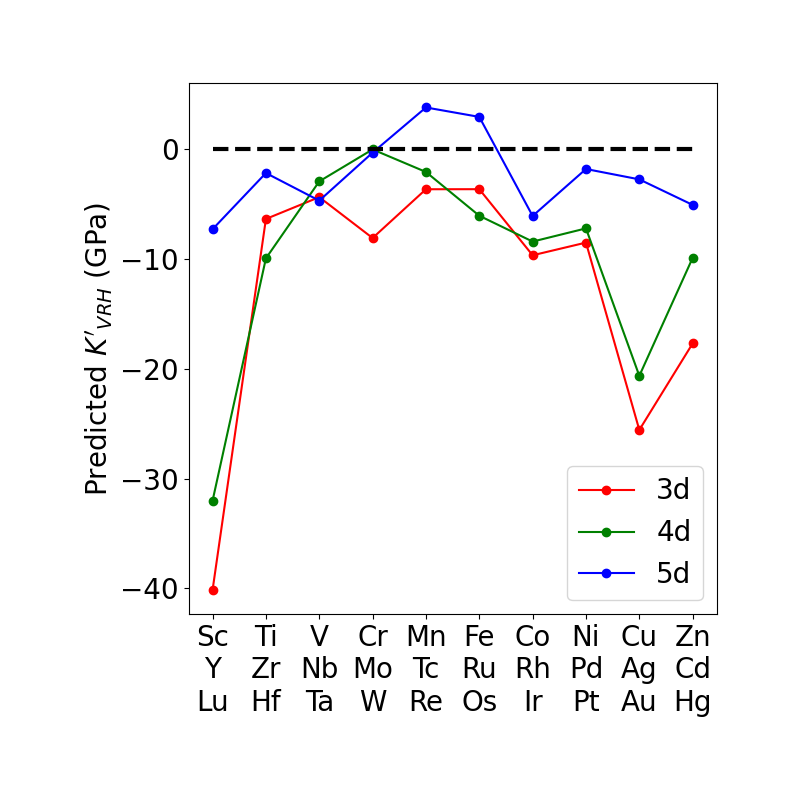}
    \caption{Predicted bulk modulus variation ($\text{K'}_{\text{VRH}}$) for a single-atom substitutionally defected Mo supercell with respect to the undefected one, along the 3d, 4d and 5d series of the periodic table. The black dashed line highlights the pure Mo case.}
    \label{fig:Kvrh_3d4d5d}
\end{figure}

Due to the strong effect induced by the Sr defect, we further focus on transition metals  to evaluate how defect-induced variations fluctutate in the compositional vicinity of the host matrix. Fig.(\ref{fig:Kvrh_3d4d5d}) shows the results along the list of 3d, 4d and 5d transition elements, where the variations are harder to distinguish from the prior periodic table plot. We can recognize a decreasing trend that is supportive of our interpretation, but in the compositional vicinity of the host matrix the fluctuations in the defect-induced variations are comparable with the MAE on bulk moduli prediction, just as the differences between the curves. 

However, in view of a trend analysis of the variations along the periodic table, Alkali (like K), Alkaline-earth (like Sr and Ba) and post-transition metals (like Se and Te), can support the given interpretation in terms of correlation with the bulk modulus of the impurity, given their associated fluctuations are much larger than the model predictions errors.

\begin{figure}[H]
    \centering
    \includegraphics[width=\textwidth]{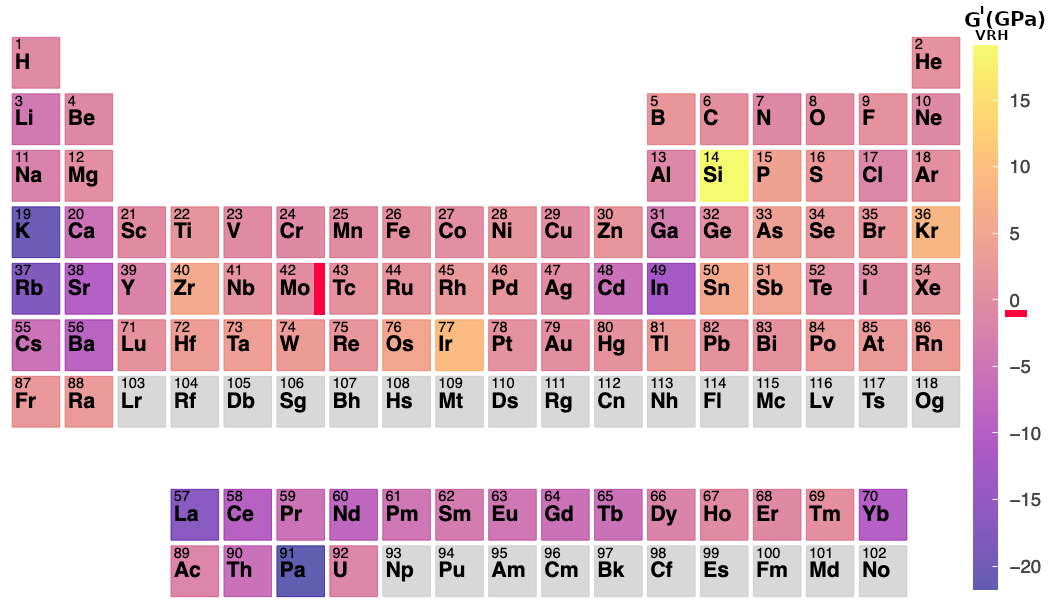}
    \caption{Predicted shear modulus variation ($\text{G'}_{\text{VRH}}$) for a single-atom substitutionally defected Mo supercell with respect to the undefected one, for each possible defect atomic specie from the provided periodic table. Similar plots for Al, Au and Ni supercells are provided in the Supplementary Materials. The red flag highlights the zero relative difference, meaning the pure Mo matrix selection.}
    \label{fig:SMpred_ptable}
\end{figure}

We may perform analogous investigations for the shear modulus of a pure host matrix, like Mo, and how it gets influenced by single-atom substitutional alloying, spanning over almost the entire periodic table, as shown in Fig.(\ref{fig:SMpred_ptable}). We follow the same protocol and  consider the variation $\text{G'}_{\text{VRH}}=\text{G}^{\text{Mo(X)}}_{\text{VRH}}-\text{G}^{\text{Mo}}_{\text{VRH}}$. In this case, the colormap reveals Si as an outlier towards hardening of the Mo matrix. This is a feature that cannot be explained by a correlation in defect- and defect-induced properties trends, since the predicted value is larger that the model prediction error on the shear modulus. We believe that the power of this investigation method is in paving the way to an efficient exploration of substitutional alloying, with the twofold possibility of looking for comparable performances (discovery of alternatives) or outstanding ones (discovery of exceptionals). Similarly to the investigation of the bulk modulus, the Alkali and Alkaline-earth metals like K, Rb, Sr and Ba are among the substitutional species providing the largest decreases in the shear modulus.

Overall, the effects and fluctuations caused by the substitutional defects on the defected host can always be highlighted, but it is not the aim of this work to find an exhaustive explanation for the existing predictions: The reasons for such trends may be due to any of the input parameters, such as the atomic number and bond lengths, or an abstract notion of local environment which is good enough to show reasonable correlations with existing alternative descriptors (i.e. atomic properties). The variations along the 3d, 4d and 5d transition metals are, for most of the cases, below the model MAE for $\text{G}_{\text{VRH}}$ predictions, therefore no meaningful extrapolation is possible, but we report the plot in Fig.(\ref{fig:g_vrh_mo_ni_bcc_orbital_sup}) of the Appendix. 

\begin{figure}[H]
    \centering
    \includegraphics[width=\textwidth]{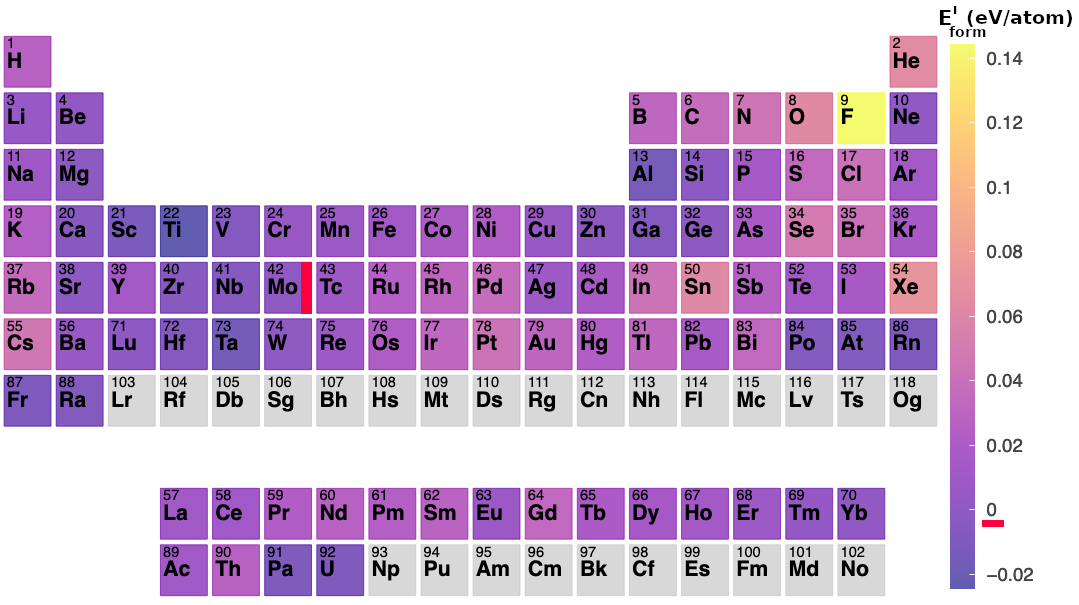}
    \caption{Predicted formation energy variation ($\text{E'}_{\text{form}}$) for a single-atom substitutionally defected Mo supercell with respect to the undefected one, for each possible defect atomic specie from the provided periodic table. Similar plots for Al, Au and Ni supercells are provided in the Supplementary Materials. The red flag highlights the zero relative difference, meaning the pure Mo matrix selection.}
    \label{fig:FEpred_ptable}
\end{figure}

\begin{figure}[H]
    \centering
    \includegraphics[width=0.65\textwidth]{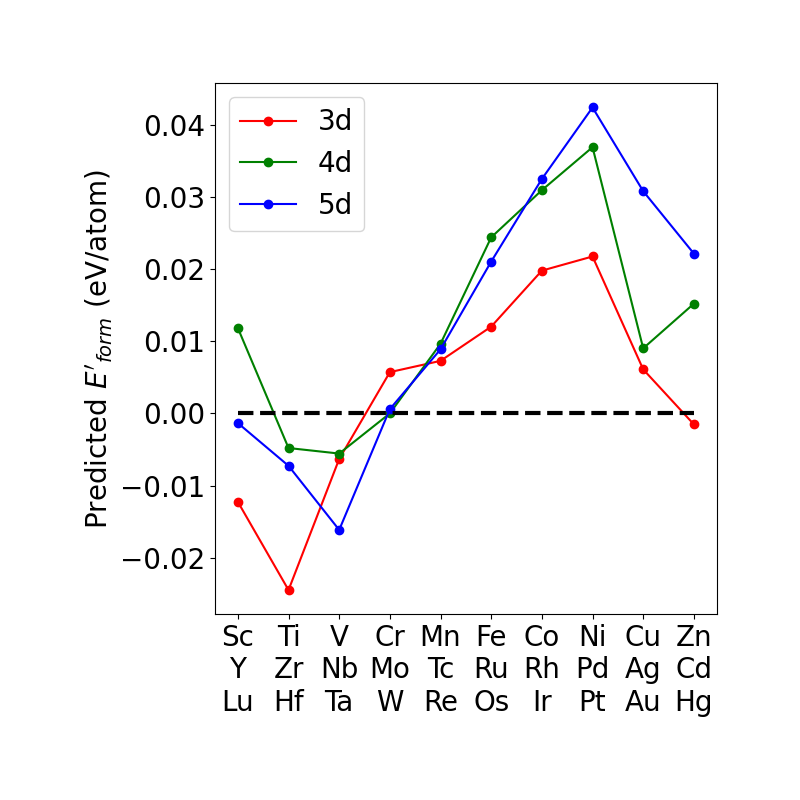}
    \caption{Predicted formation energy variation ($\text{E'}_{\text{form}}$) for a single-atom substitutionally defected Mo supercell with respect to the undefected one, along the 3d, 4d and 5d series of the periodic table. The black dashed line highlights the pure Mo case.}
    \label{fig:Ef_3d4d5d}
\end{figure}

For what concerns the formation energy, by definition of the latter, the results in Fig.(\ref{fig:FEpred_ptable}) can be interpreted as the gain or loss in stability after the single-atom substitutional alloying has taken place. Fluorine represents a strong outlier, raising the formation energy by 0.14 eV/atom with respect to the pure Mo matrix. As shown in Fig.(\ref{fig:Ef_3d4d5d}), it is evident that there is a trend that spans over the periods, and an overall interesting correlation could be found with the known trends for the atomic electronegativity along the periodic table, suggesting that the higher the latter for the substitutional defect in the Mo matrix, the higher is the resulting formation energy. 

\subsection{Property prediction - DFT validation}\label{subsec:val_results}

The validation of artificial intelligence methods predictions is a crucial step in order to quantify their quality. Even though the proposed graph network based method has been already validated for its predictions in bulk crystals, this work aims at testing its capabilities in the presence of single-atom substitutional defects. As explained in Sec.\ref{subsec:validation}, we compare the model predictions of $\text{K}_{\text{VRH}}$, $\text{G}_{\text{VRH}}$ and $\text{E}_{\text{form}}$ with the DFT based ones, obtained with the \texttt{THERMO\_PW} driver of \texttt{QE}. Table(\ref{table:valid}) shows the results for the validation on the properties in the case of an Al matrix and single-atom substitutional defects, including B, C, I, Ni and Zr. 

\begin{table}[htb]
\centering
\begin{tabular}{ | l | l | r | r | r |}
    \hline
   System & Method & $\text{K}_{\text{VRH}}$ (GPa) & $\text{G}_{\text{VRH}}$ (GPa) & $\text{E}_{\text{form}}$ (eV/atom)  \\ \hline \hline
   \multirow{2}{*}{$\text{Al}_{\text{B}}$} &MEGNet &84.456&60.405&0.050 \\ 
                         &DFT &78.101& 24.074&0.096 \\ \hline
   \multirow{2}{*}{$\text{Al}_{\text{C}}$} &MEGNet &87.165&34.762&0.113 \\ 
                         &DFT &73.791&16.673&0.228 \\ \hline
   \multirow{2}{*}{$\text{Al}_{\text{I}}$} &MEGNet &72.697&29.796&0.056 \\ 
                         &DFT &63.655&13.514&0.475 \\ \hline
   \multirow{2}{*}{$\text{Al}_{\text{Ni}}$} &MEGNet &84.683&30.437&0.009 \\ 
                         &DFT & 81.363&37.086&4.567 \\ \hline
   \multirow{2}{*}{$\text{Al}_{\text{Zr}}$} &MEGNet &87.881&29.158&-0.048 \\ 
                         &DFT &79.672& 30.068&1.262 \\ \hline
   MAE &  & 8.060 & 15.653 & 1.290 \\ \hline
   \end{tabular}
    \caption{Comparison of the DFT and MEGNet results for the three properties of interest in a small set of samples. Here, $\text{Al}_{\text{B}}$ means a single B-atom substitution in an Al host $2\times2\times2$ supercell matrix.}
    \label{table:valid}
\end{table}
Comparing the MAE values of our DFT calculations, with the ones of the model for non-defected systems in Table(\ref{table:tableparity}), we find good performances of the model with respect to $\text{K}_{\text{VRH}}$ and $\text{G}_{\text{VRH}}$ predictions, but large errors when it comes to the formation energies: the first may be regarded as a success, given that the model is predicting properties of a new class of systems, and given the computational limitations; the second one is a negative signature, even though the (i) defect dependent nature of the error order of magnitude opens up a deeper window of investigation on its reasons, and (ii) the validation set for defected systems is extremely small compared to the undefected MP dataset, on which initial MAEs have been evaluated. 



\subsection{Size effects }\label{subsec:size_effects}

\begin{figure}[H]
    \centering
    \includegraphics[width=0.65\textwidth]{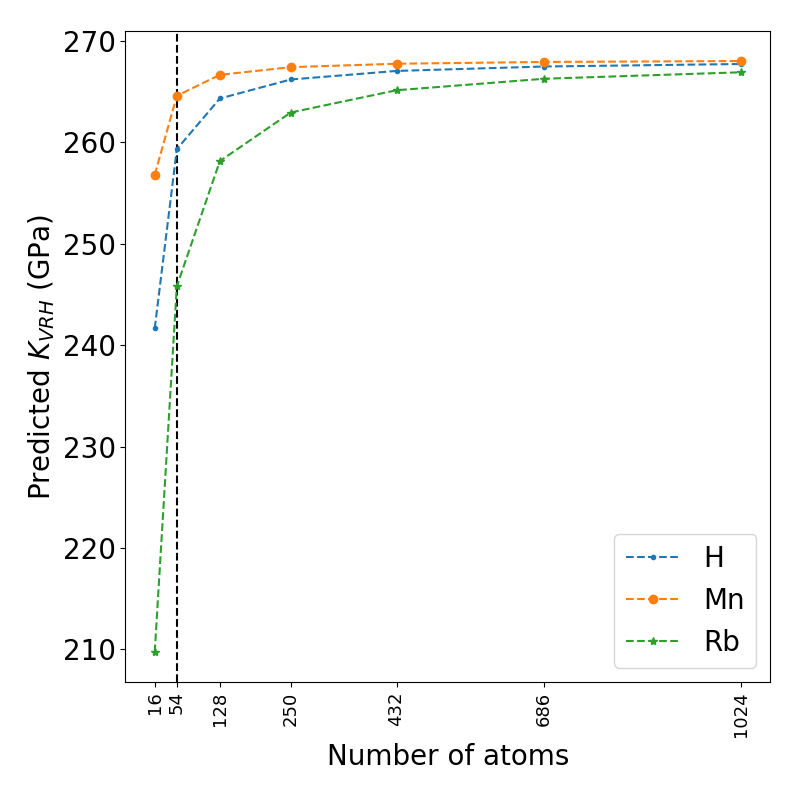}
    \caption{Saturation plot of a Mo host matrix $\text{K}_{\text{VRH}}$ when substitutionally defected with H, Mn or Rb, for different supercell sizes.}
    \label{fig:Sat1}
\end{figure}

\begin{figure}[H]
    \centering
    \includegraphics[width=0.65\textwidth]{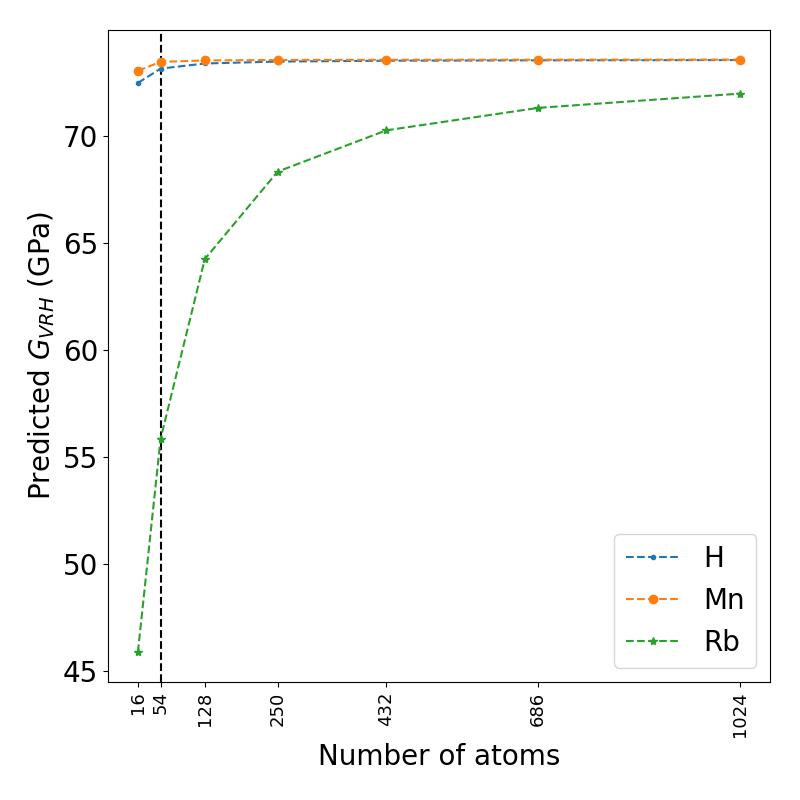}
    \caption{Saturation plot of a Mo host matrix $\text{G}_{\text{VRH}}$ when substitutionally defected with H, Mn or Rb, for different supercell sizes.}
    \label{fig:Sat2}
\end{figure}

\begin{figure}[H]
    \centering
    \includegraphics[width=0.65\textwidth]{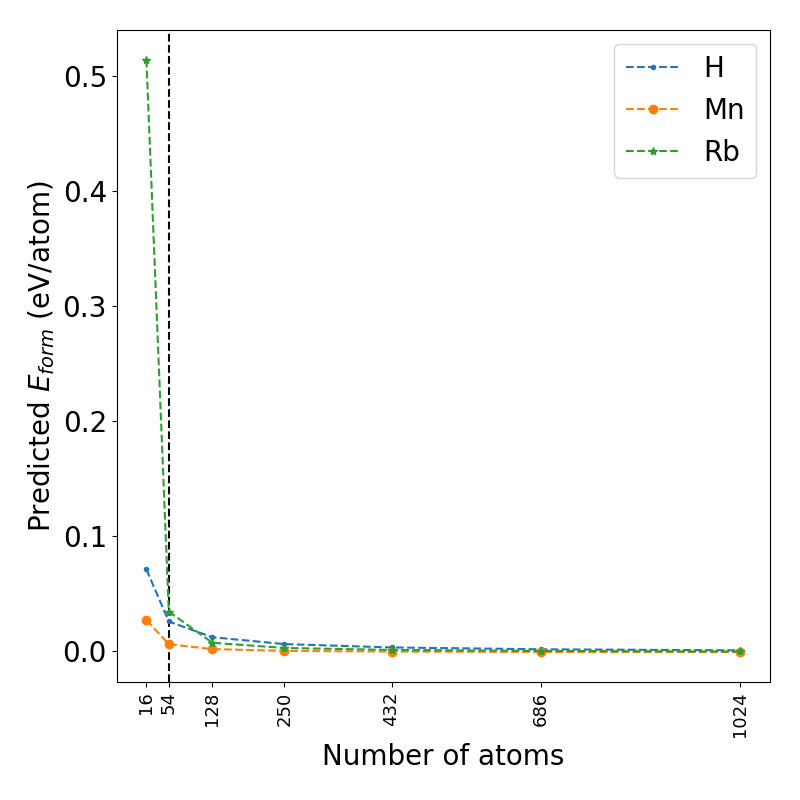}
    \caption{Saturation plot of a Mo host matrix $\text{E}_{\text{form}}$ when substitutionally defected with H, Mn or Rb, for different supercell sizes.}
    \label{fig:Sat3}
\end{figure}

{In substitutional alloying, the defect atomic specie is usually present in a dilute concentration, in the range of $0.1\text{-}10\%$. For this reason, it is of interest the study of how the property predictions vary with defect concentration, that we propose in the saturation plots of Fig.(\ref{fig:Sat1}), Fig.(\ref{fig:Sat2}) and Fig.(\ref{fig:Sat3}). Our example system follows the choice of the Mo host matrix, with the H, Mn and Rb substitutional defects we focused on, respectively in the second and first part of the previously reported results. Interstingly, a hierarchy is conserved among the defect-induced variations for different host supercell sizes: the single Rb defect always causes the largest deviations of the studied properties from their asymptotic over-dilute level ($<0.1\%$). Moreover, while the defect formation energy, as expected, shows a common descent to zero-level for all the defect species, the predicted structural properties seem to be sensitive on them, with the Rb-defected Mo crystal conserving a visible difference in the property value even at the limit of $0.1\%$ concentration, both for $\text{K}_{\text{VRH}}$ and $\text{G}_{\text{VRH}}$. Even though it is out of the present work aims to validate the asymptotic-size behaviour of the predictions with accurate but expensive DFT calculations, we believe these results to stand in favour of the positive model understanding of a crystalline defected system environment.
}  

\begin{figure}[H]
    \centering
    \includegraphics[width=0.95\textwidth]{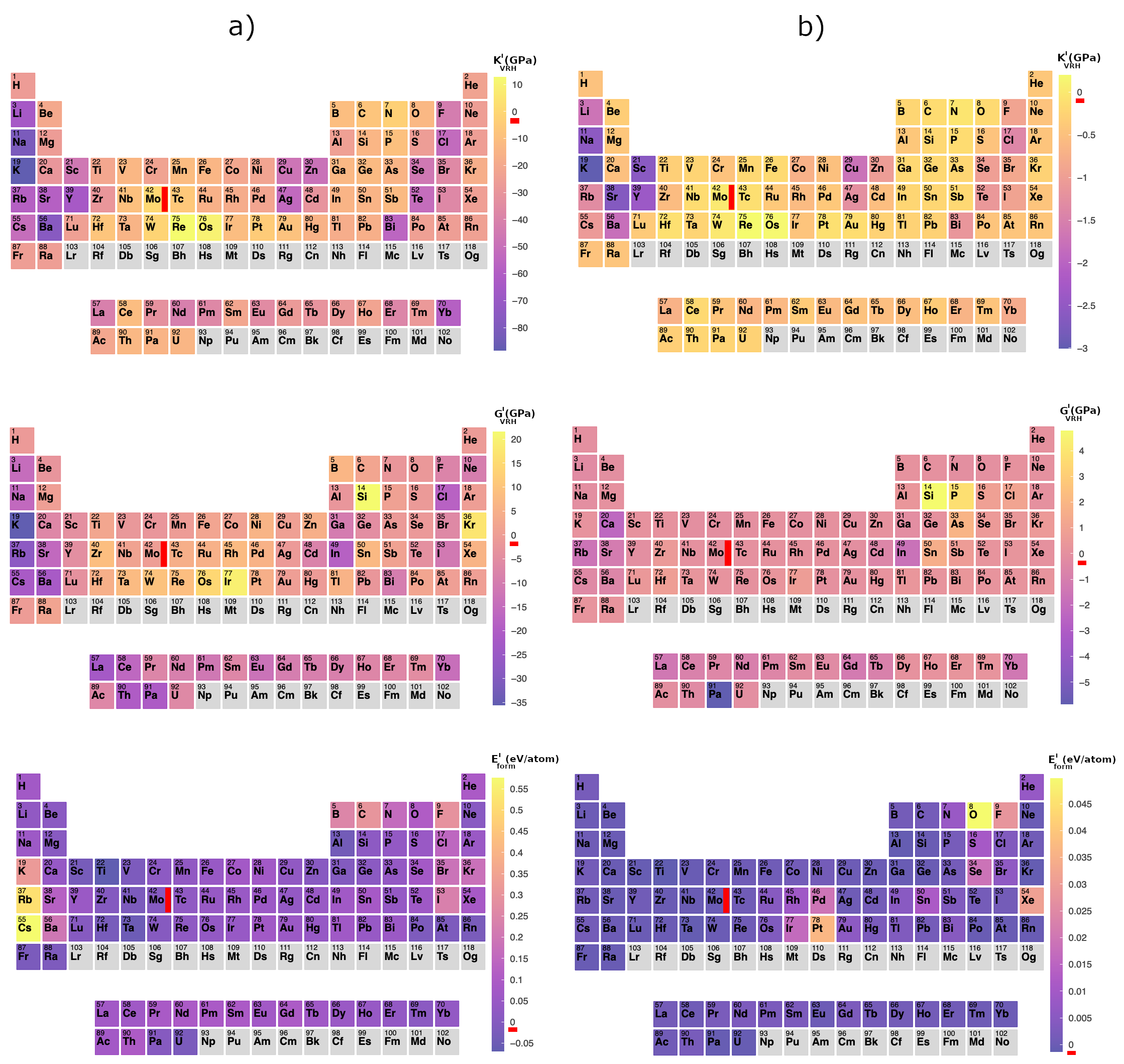}
    \caption{Predicted variations $\text{K'}_{\text{VRH}}$ (first row plots), $\text{G'}_{\text{VRH}}$ (second row plots), $\text{E'}_{\text{form}}$ (third row plots) for a single-atom substitutionally defected Mo supercell with respect to the undefected one, for each possible defect atomic specie from the provided periodic table. In plots of column a) a $2\times2\times2$ supercell is considered, while in column b) a $8\times8\times8$ supercell.}
    \label{fig:Sat_ptables}
\end{figure}

{Let us continue along the previously paved path of analysis which also looks into the effects of systematically changing the substitutional atomic specie among the large variety contained in the periodic table, as previously done in Fig.(\ref{fig:BMpred_ptable}), with the only exception of selecting the smallest and largest supercell sizes from the previous saturation plots of Fig.(\ref{fig:Sat1}), respectively $2\times2\times2$ ($1\%$ concentration, $16$ atoms cell) and $8\times8\times8$ ($10\%$ concentration, 1024 atoms cell). In Fig.(\ref{fig:Sat_ptables}), focusing on the first row of plots which deal with the $\text{K'}_{\text{VRH}}$ variation, and comparing with the previously investigated supercell case of size $3\times3\times3$ ($2\%$ concentration) of Fig.(\ref{fig:BMpred_ptable}), one can notice that the scale of property variations changes accordingly: smaller defect concentrations lead to smaller defect-induced effects, and viceversa, as expected. In particular, in the largest supercell case the induced variations range is reduced to a $3\%$ of the $3\times3\times3$ system's one. However, the composition map outliers are left unchanged. The shear modulus variations, in the second row of tables, sees the emergence of new outliers in the chemical neighbourhoods of the previously obtained ones in $3\times3\times3$ supercells, while the formation energy variations undergo both a change in scale and a complete change in the map outliers. Following the brief assessment of predictions quality performed for each of the interesting properties in the previous section, we expect the formation energy variations to suffer by non-negligible fluctuations, leading to the possible need of further validation. However, the main aim of the present discussion is to underline the power of the overall approach in highlighting the path towards composition search in substitutional alloying, which can effectively drive to specific desired emerging properties.}

\backmatter

\section{Conclusions}

In this work, we utilized a convolutional graph-neural network, based on the MEGNet architecture~\cite{megnetmodel}, in order to attempt the design of novel alloys. Alloying involves substitutional and interstitial alloying at relatively low concentrations, thus, single-defect properties shall be informative on overall designing capabilities and guidance. For this purpose, we utilized the MP database, and we focused on the prediction of the properties of single-atom substitutionally defected bulk crystals,in the context of both (i) systematic substitution with a specific set of species in a wide variety of crystals from the entire Materials Project dataset and (ii) systematic substitution with a variety of atomic species in a specific set of pure bulk crystals. We also validated some of the results with our own DFT calculations. We believe that such approaches might provide novel insights into alloy design, especially if predictions include extended lattice defects such as dislocations and/or grain boundaries.

\bmhead{Acknowledgments}
We would like to thank J. Llorca for insights and fruitful discussions and suggestions.  This research was funded in part by the European Union Horizon 2020 research and innovation program under Grant Agreement No. 857470 and from the European Regional Development Fund via Foundation for Polish Science International Research Agenda PLUS program Grant No. MAB PLUS/2018/8.

\subsection{Data availability}
Requests for additional data and availability can be directed to the provided lead contact. The employed DFT data, ML model and database can be found on the GitHub: https://github.com/danielcieslinski/suballoy

\section*{Declarations}
Declarations






\begin{appendices}

\section{Further Results for: Property prediction - A defect in a wide range of matrices }\label{secA1}
\begin{figure}[H]
\centering
\begin{subfigure}{.5\textwidth}
  \centering
  \includegraphics[width=.99\linewidth]{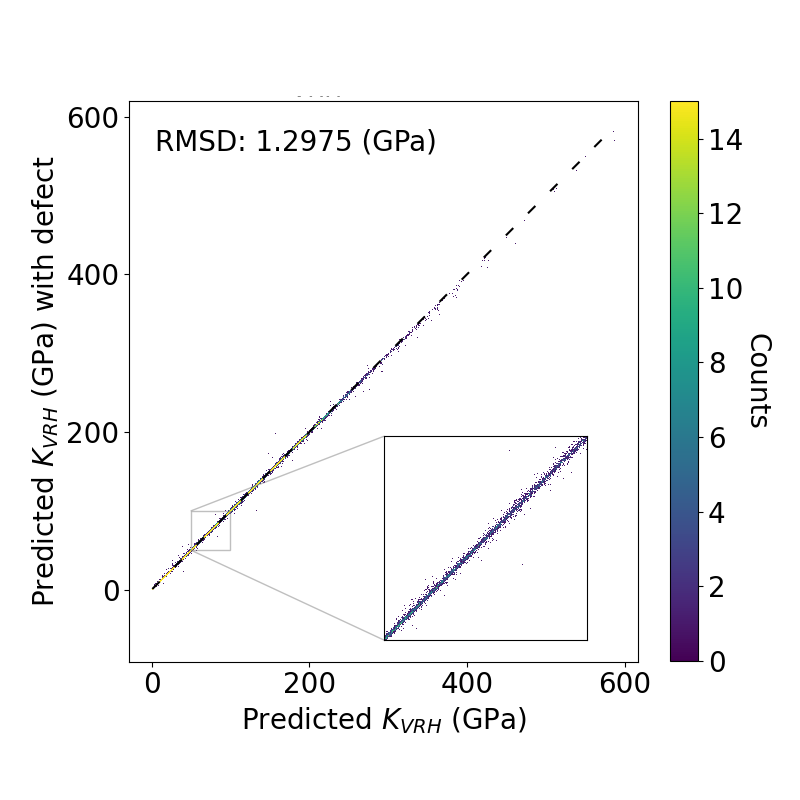}
  \caption{H}
  \label{fig:kvrh_H_defect_3_sup}
\end{subfigure}%
\begin{subfigure}{.5\textwidth}
  \centering
  \includegraphics[width=.99\linewidth]{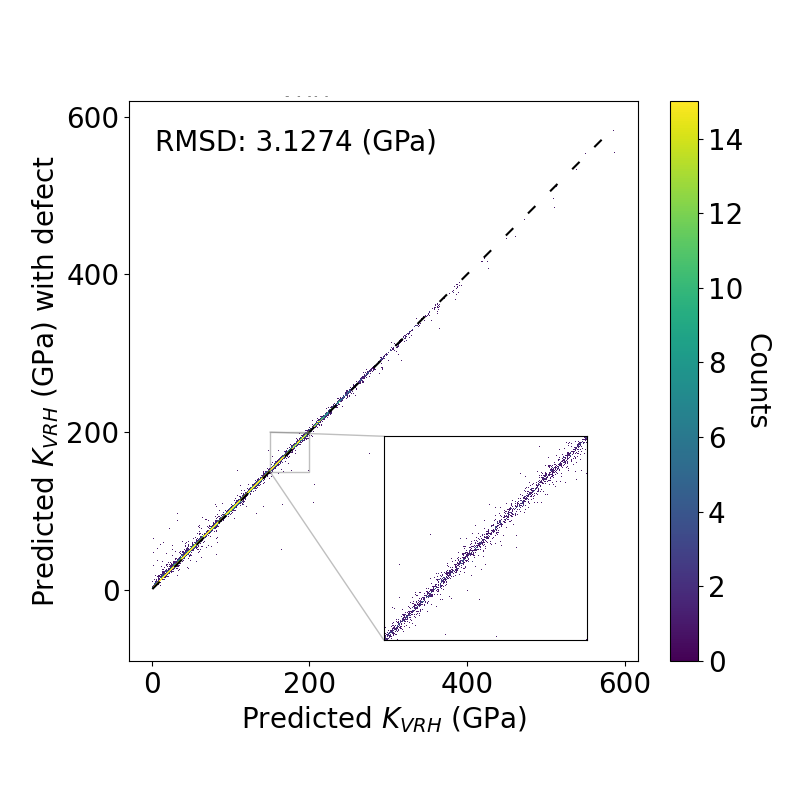}
  \caption{Mn}
  \label{fig:kvrh_Mn_defect_3_sup}
\end{subfigure}
\caption{Predicted bulk modulus in the H- and Mn-defected MP crystals with respect to its prediction in non-defected ones.}
\label{fig:kvrh_H_Mn_defect_3_sup}
\end{figure}

\begin{figure}[H]
\centering
\begin{subfigure}{.5\textwidth}
  \centering
  \includegraphics[width=.99\linewidth]{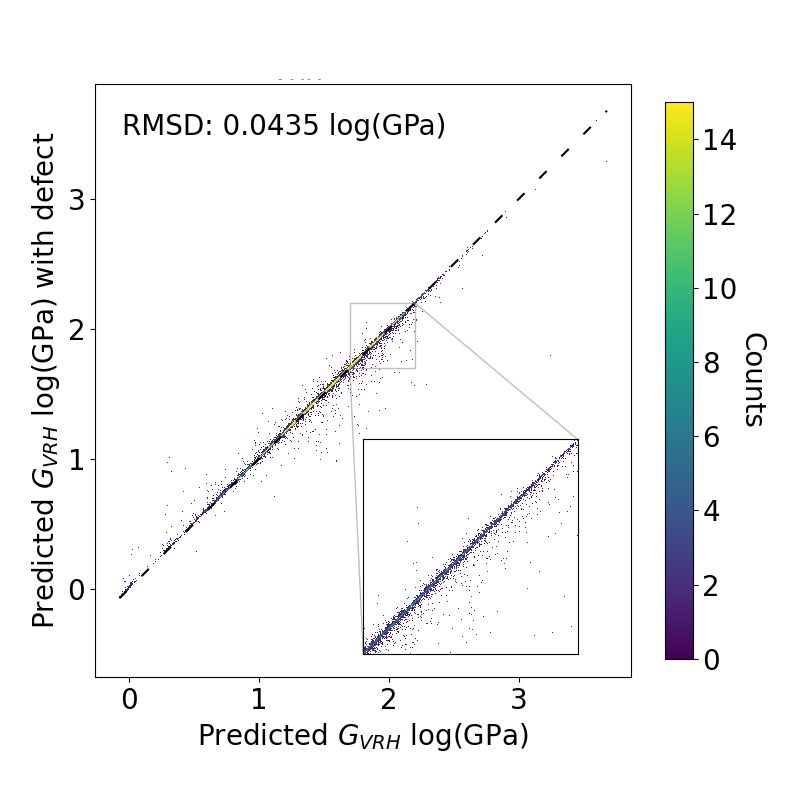}
  \caption{H}
  \label{fig:gvrh_H_defect_3_sup}
\end{subfigure}%
\begin{subfigure}{.5\textwidth}
  \centering
  \includegraphics[width=.99\linewidth]{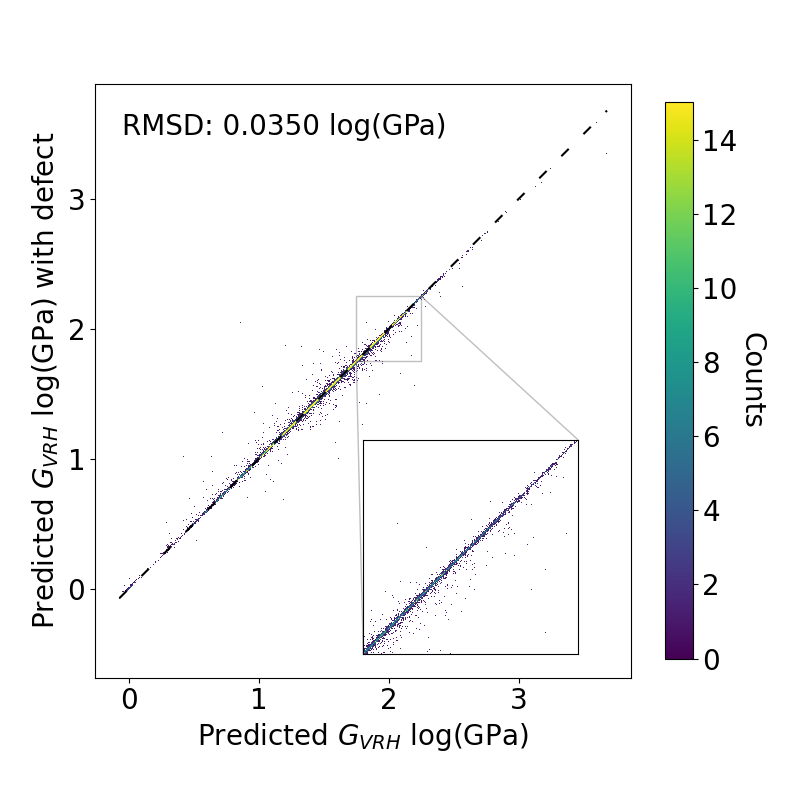}
  \caption{Mn}
  \label{fig:gvrh_Mn_defect_3_sup}
\end{subfigure}
\caption{Predicted shear modulus in the H- and Mn-defected MP crystals with respect to its prediction in non-defected ones.}
\label{fig:gvrh_H_Mn_defect_3}
\end{figure}

\begin{figure}[H]
\centering
\begin{subfigure}{.5\textwidth}
  \centering
  \includegraphics[width=.99\linewidth]{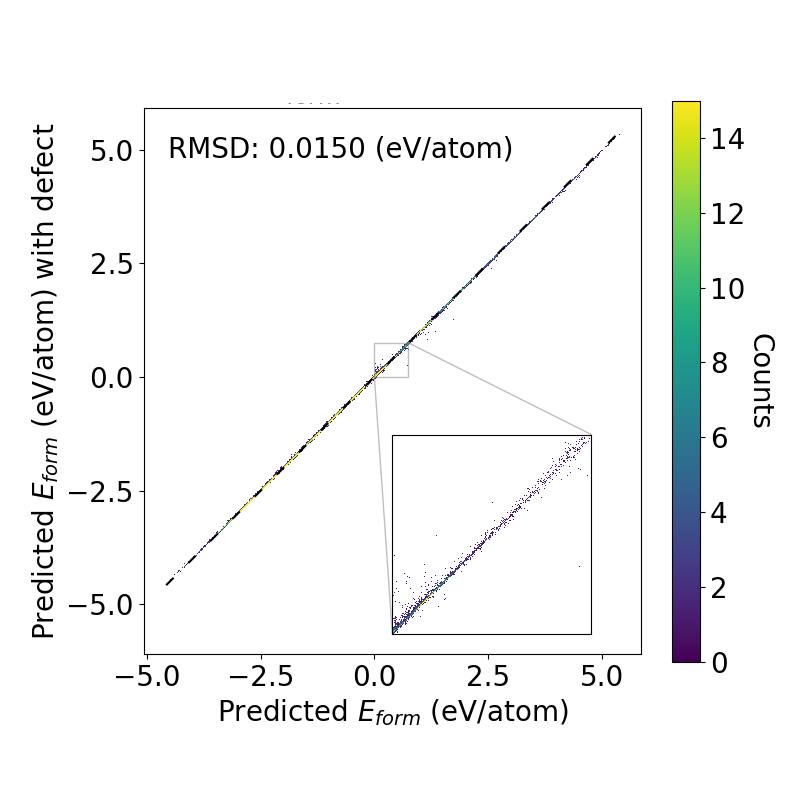}
  \caption{Mn}
  \label{fig:eform_Mn_defect_3_sup}
\end{subfigure}%
\begin{subfigure}{.5\textwidth}
  \centering
  \includegraphics[width=.99\linewidth]{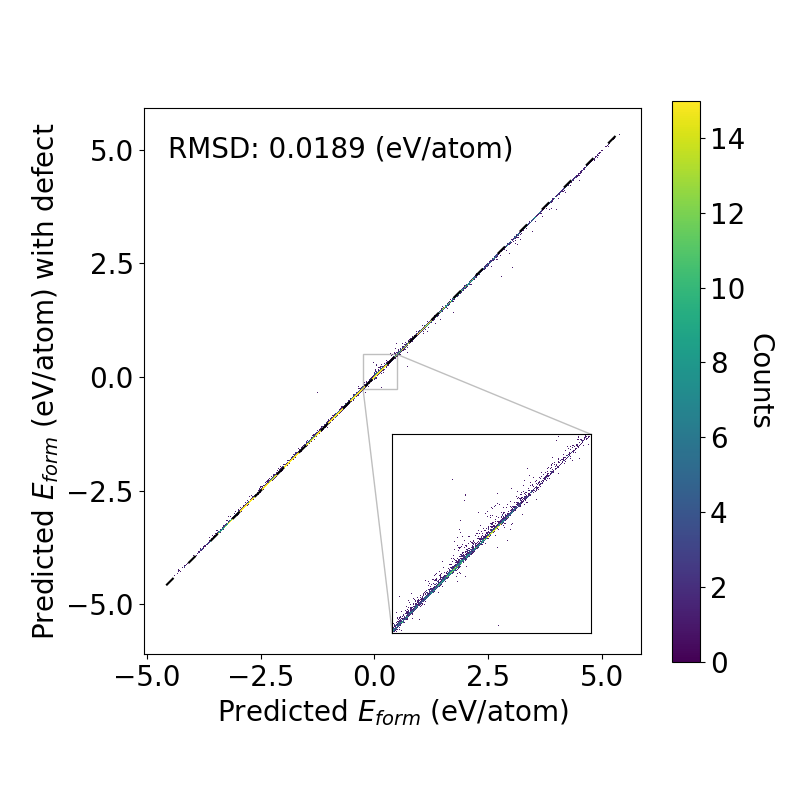}
  \caption{Rb}
  \label{fig:eform_Rb_defect_3_sup}
\end{subfigure}
\caption{Predicted formation energy in the Mn- and Rb-defected MP crystals with respect to its prediction in non-defected ones.}
\label{fig:eform_Mn_Rb_defect_3}
\end{figure}

As mentioned in Sec.(\ref{subsec:Defectsinamatrix}), we first focus on the prediction of defected systems properties, when the systematic substitutional defective process is applied with the same atom on the
Materials Project crystals dataset. There, we show the RMSD for the predictions of bulk modulus, shear modulus, and defect formation energy, respectively in Fig.(\ref{fig:BMpred-1case}) and Table(\ref{table:BMpred-1case}), Fig.(\ref{fig:SMpred-1case}) and Table(\ref{table:SMpred-1case}), and Fig.(\ref{fig:FEpred}) and Table(\ref{table:FEpred-1case}). However, the figures are there reported only for the defects leading to the corresponding largest RMSD values. Here, for completeness, we report the missing ones in Fig.(\ref{fig:kvrh_H_Mn_defect_3_sup}), Fig(\ref{fig:gvrh_H_Mn_defect_3}) and Fig(\ref{fig:eform_Mn_Rb_defect_3}).

\begin{figure}[H]
    \centering
    \includegraphics[width=0.6\textwidth]{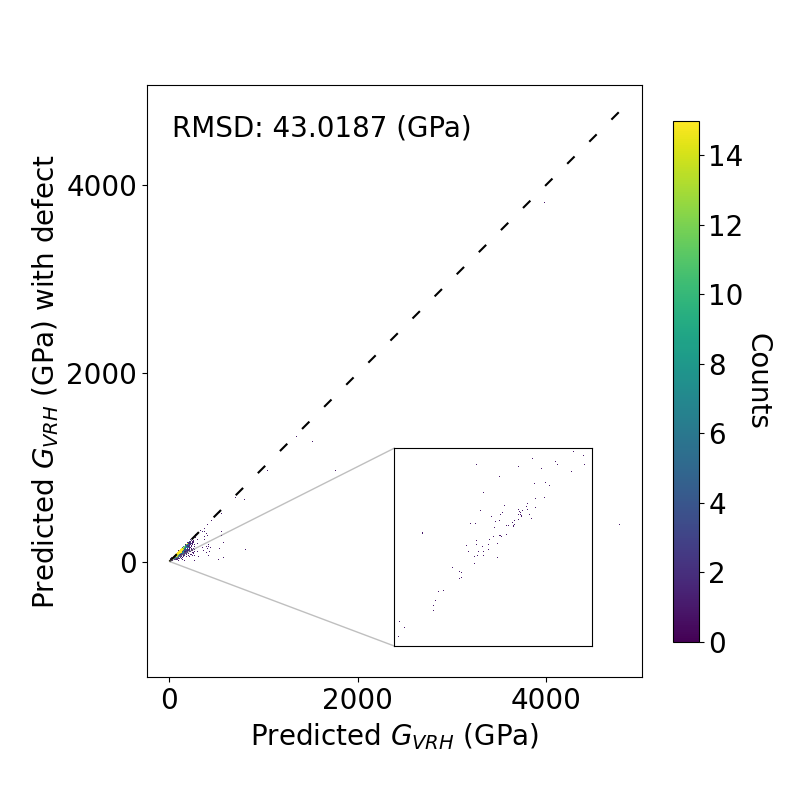}
    \caption{Predicted formation energy in the Rb-defected MP crystals with respect to its prediction in non-defected ones. The plot has the aim to display the need of a log-log scale to appreciate the deviations.}
    \label{fig:supp_gvrh_Rb_defect_3_NOLOG}
\end{figure}

For the shear modulus predictions, the reader might have noticed the choice of a different scale for the plots. A part from noticing a common use in the literature of the log-log scale for this quantity and adopting the same for the sake of comparison, we also decided to understand the reason behind this choice by plotting in a normal scale as in Fig.(\ref{fig:supp_gvrh_Rb_defect_3_NOLOG}): the wide scale over which few predictions are spreading does not allow to appreciate the distribution of the data. 

\section{Further Results for: Property prediction - A wide range of defects in the same matrix }\label{secA2}
\begin{figure}[H]
    \centering
    \includegraphics[width=\textwidth]{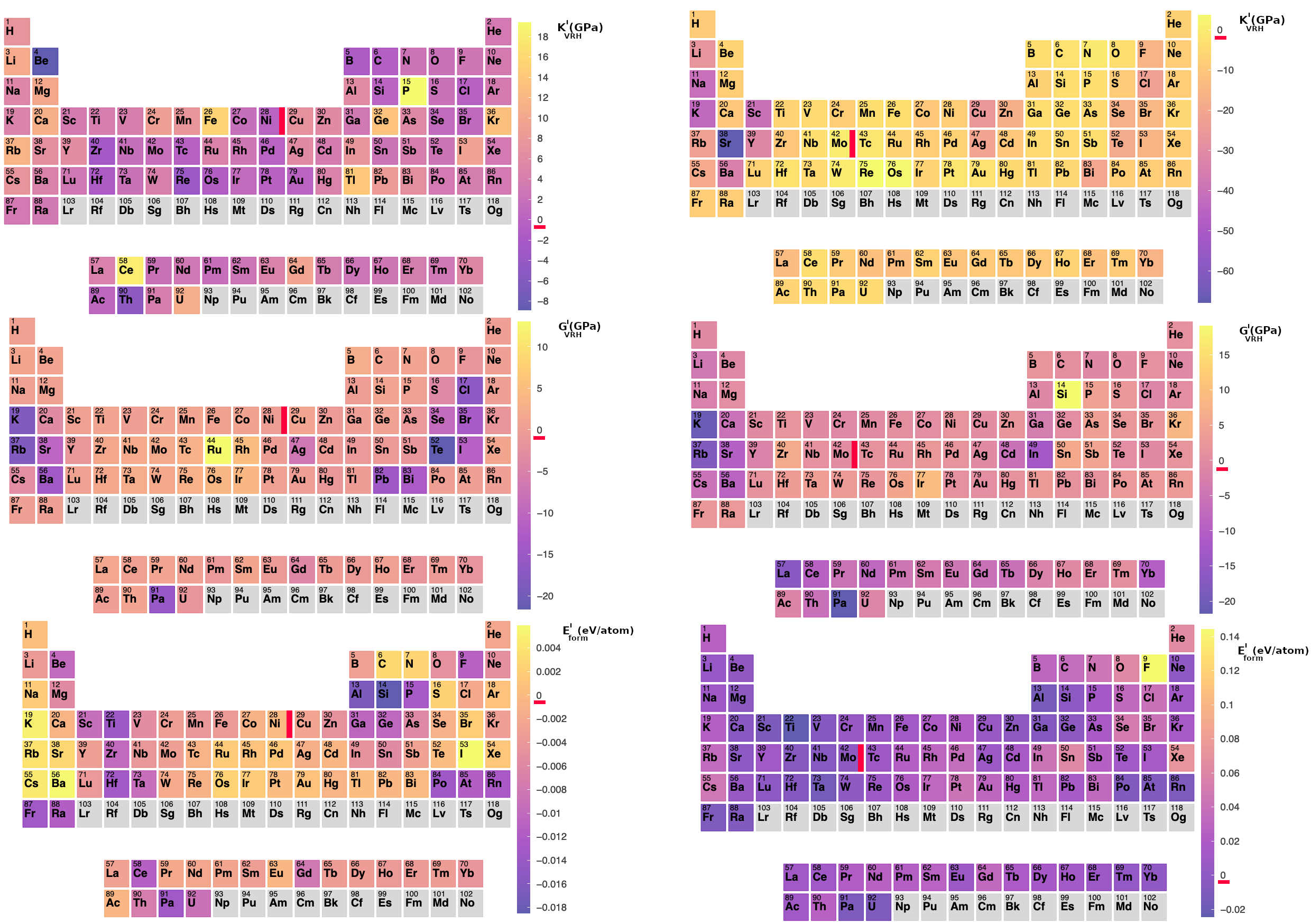}
    \caption{Comparison between the periodic table plots of the predicted bulk modulus, shear modulus and formation energy for a single-atom substitutionally defected Ni (left column) and Mo (right column) supercell with respect to the undefected one, for each possible atomic specie from the provided periodic table.}
    \label{fig:supp_combined1}
\end{figure}

\begin{figure}[H]
    \centering
    \includegraphics[width=\textwidth]{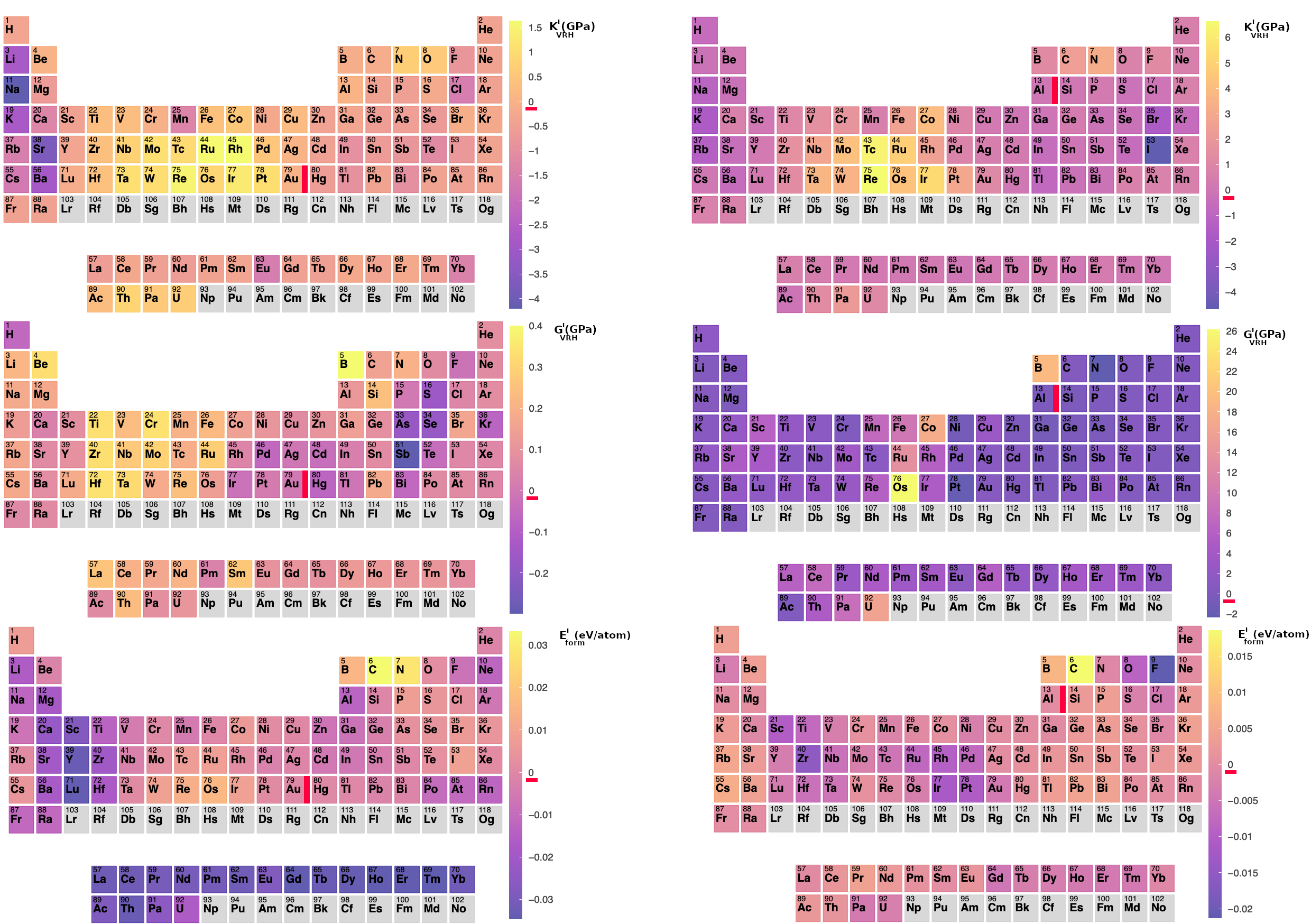}
    \caption{Comparison between the periodic table plots of the predicted bulk modulus, shear modulus and formation energy for a single-atom substitutionally defected Au (left column) and Al (right column) supercell with respect to the undefected one, for each possible atomic specie from the provided periodic table.}
    \label{fig:supp_combined2}
\end{figure}

Following a similar selection criterion for the results to which the main body of the manuscript is dedicated, for the second part of our investigation, in which we consider the systematically change of the single-atom substitutional defect specie in the same crystal structure, we decided to show the results for the Molybdenum host matrix, which was displaying, overall, the largest variations in the interesting quantities. In Fig.(\ref{fig:supp_combined1}) and Fig.(\ref{fig:supp_combined2}), we report the collection of defect-induced variations in the bulk modulus, shear modulus, and formation energy for the all host-matrices investigated here, respectively Ni and Mo, and Au and Al. A composite visualization of all the variational periodic tables allows for an overall comparison, and helps in appreciating that interesting effects in the previously not shown matrices are not missing: \begin{itemize}
    \item the shear modulus variation scale of the Ni matrix upon single-atom substitution is comparable to the one of Mo, and they also share alkaline and alkaline earth metals in the lower bound variations;
    \item similarly to the previous point, Au and Al share a similar variation scale and outlier map for the bulk modulus, and, in particular, for the formation energy
\end{itemize}

Even though the variations in the interesting properties upon substitutional-defecting of the host matrices are dominant for the Mo matrix, we here want to report, as previously done, all the variations, and draw a brief comparison between them in a combined visualization. While the trends in the bulk modulus variations of Mo and Ni matrices are noticeably different, Al and Au show pretty similar trends and hierarchies on different scales: the central 4d and 5d elements tend to maximize the variations, with a minimization happening at extremas and, for both, for a Mn substitution. In particular, it is interesting to notice that in the Al matrix nearly all the periodic table elements considered lead to a positive bulk modulus variation. A similarity between variations in Mo and Ni holds for the shear modulus variations, which share comparable scales and a similar exchanged role of 5d (in Mo) and 4d (in Ni) substitutions, and for the formation energy variations, showing remarkably similar trends over scales differing by an order of magnitude. 

\begin{figure}[H]
\centering
\begin{subfigure}{.5\textwidth}
  \centering
  \includegraphics[width=.99\linewidth]{k_vrh_mo_bcc_orbital.png}
  \caption{Mo}
  \label{fig:k_vrh_mo_bcc_orbital_sup}
\end{subfigure}%
\begin{subfigure}{.5\textwidth}
  \centering
  \includegraphics[width=.99\linewidth]{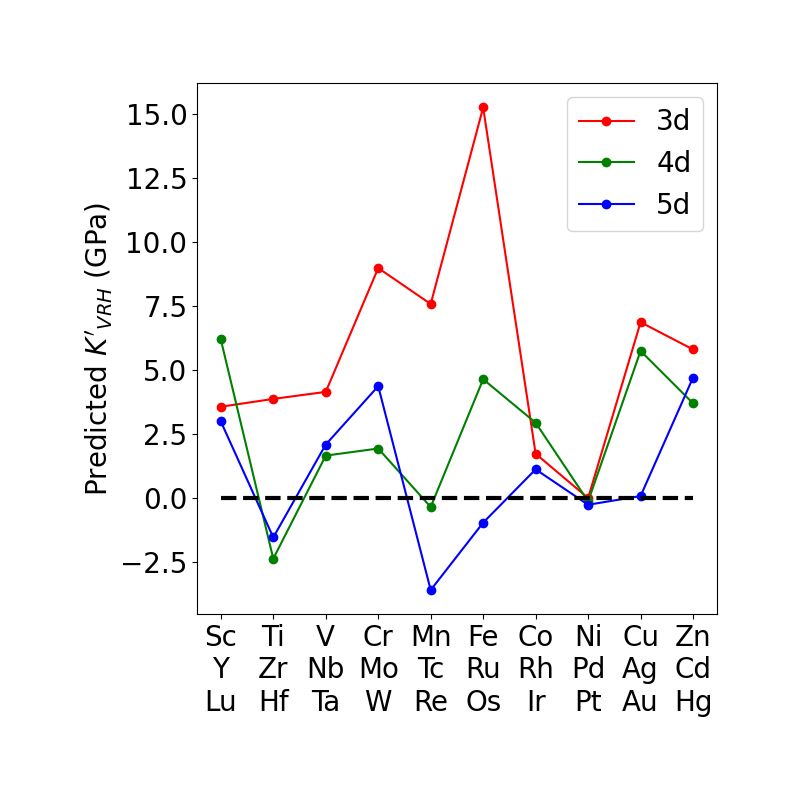}
  \caption{Ni}
  \label{fig:k_vrh_ni_fcc_orbital_sup}
\end{subfigure}
\caption{Predicted bulk modulus variation ($\text{K'}_{\text{VRH}}$) for a single-atom substitutionally defected Mo and Ni supercell with respect to the undefected one, along the 3d, 4d and 5d series of the periodic table. The black dashed line highlights the pure host matrix case.}
\label{fig:k_vrh_mo_ni_fcc_orbital}
\end{figure}

\begin{figure}[H]
\centering
\begin{subfigure}{.5\textwidth}
  \centering
  \includegraphics[width=.99\linewidth]{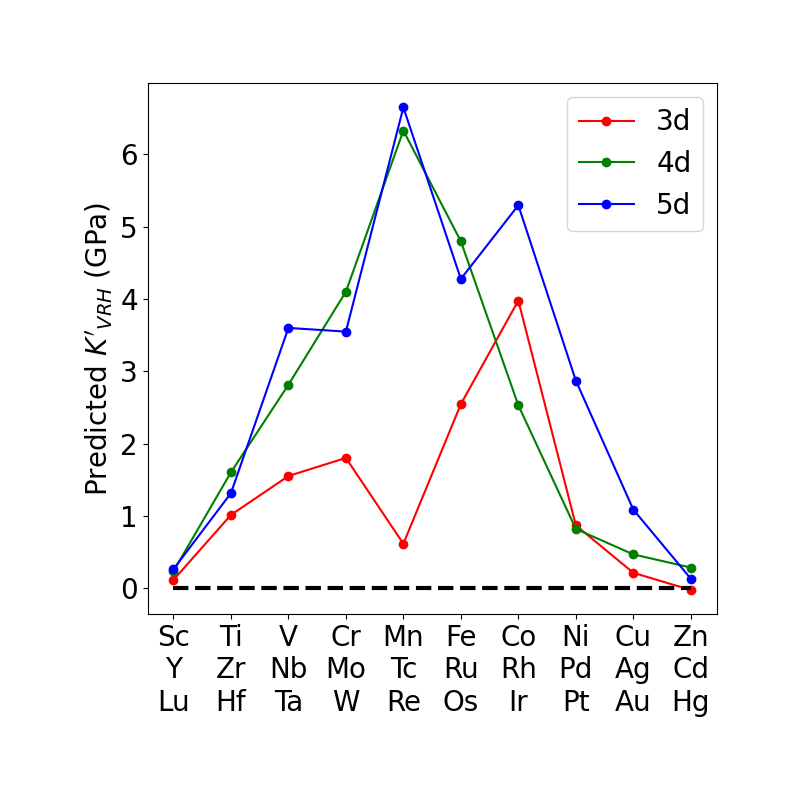}
  \caption{Al}
  \label{fig:k_vrh_al_fcc_orbital_sup}
\end{subfigure}%
\begin{subfigure}{.5\textwidth}
  \centering
  \includegraphics[width=.99\linewidth]{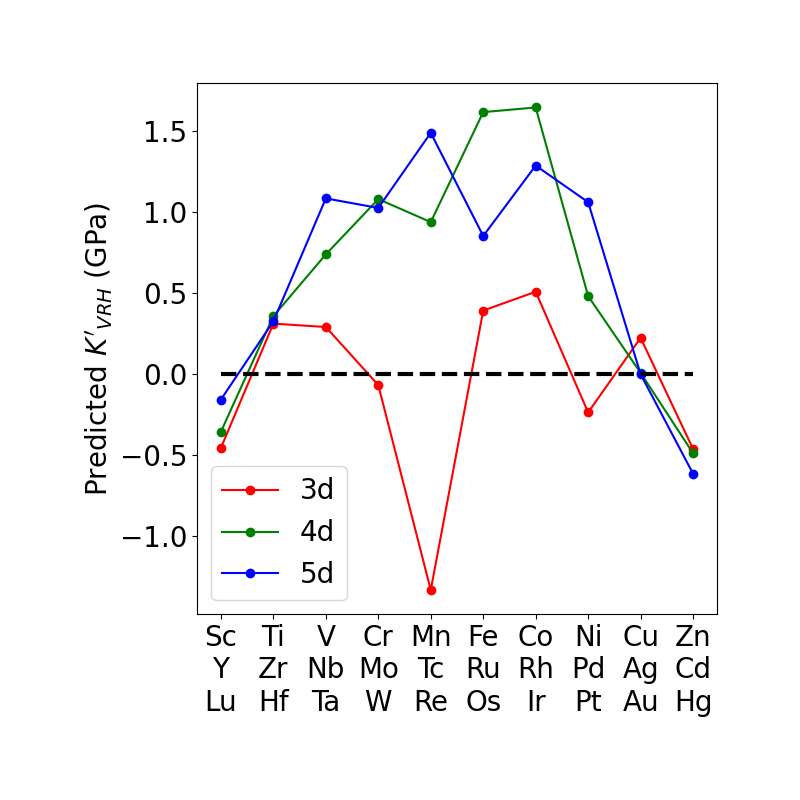}
  \caption{Au}
  \label{fig:k_vrh_au_fcc_orbital_sup}
\end{subfigure}
\caption{Predicted bulk modulus variation ($\text{K'}_{\text{VRH}}$) for a single-atom substitutionally defected Al and Au supercell with respect to the undefected one, along the 3d, 4d and 5d series of the periodic table. The black dashed line highlights the pure host matrix case.}
\label{fig:k_vrh_al_au_fcc_orbital_sup}
\end{figure}

\begin{figure}[H]
\centering
\begin{subfigure}{.5\textwidth}
  \centering
  \includegraphics[width=.99\linewidth]{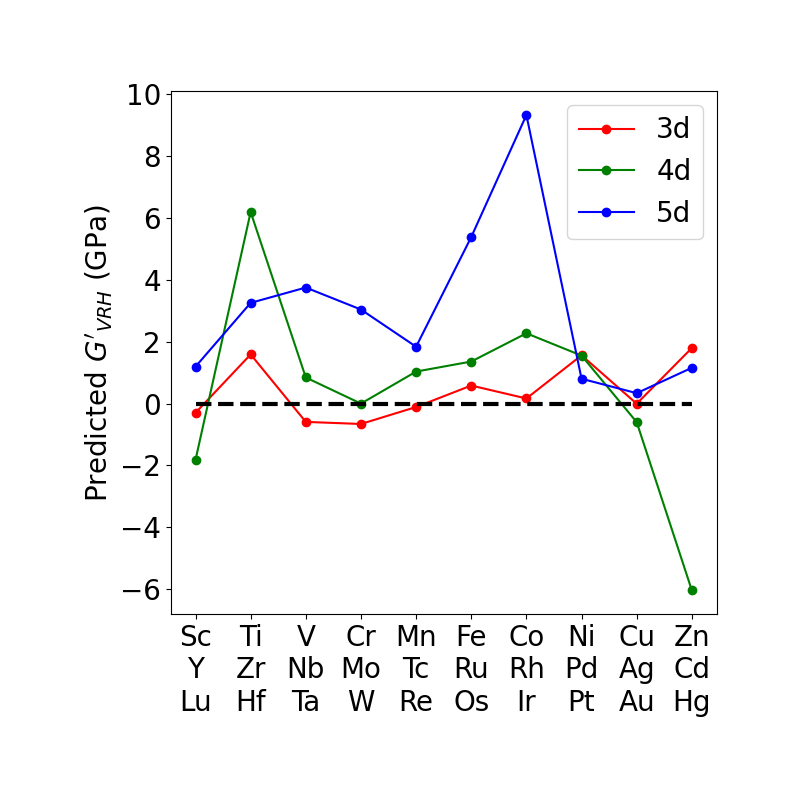}
  \caption{Mo}
  \label{fig:g_vrh_mo_bcc_orbital_sup}
\end{subfigure}%
\begin{subfigure}{.5\textwidth}
  \centering
  \includegraphics[width=.99\linewidth]{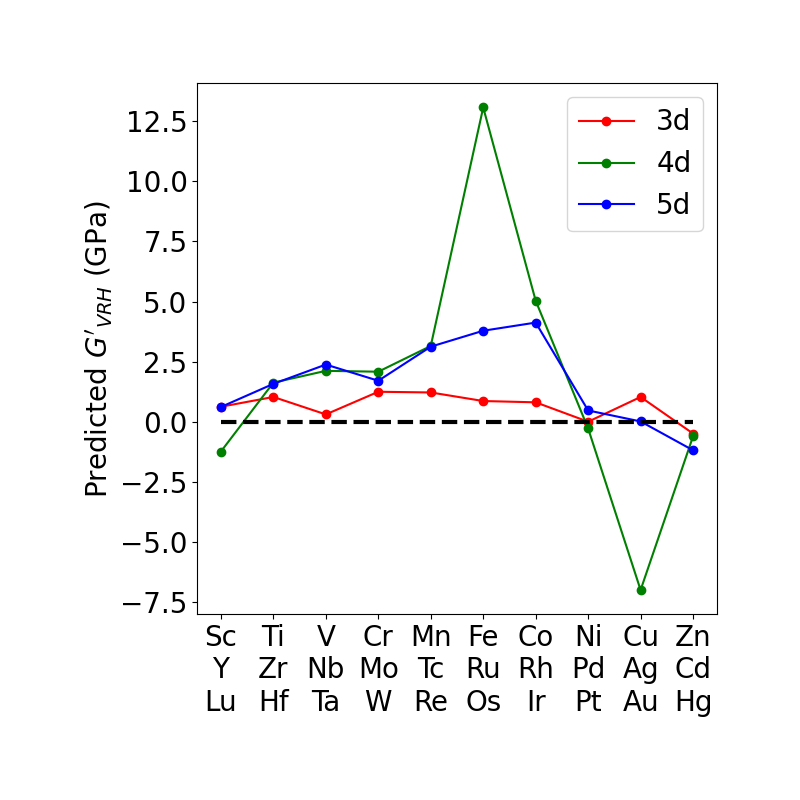}
  \caption{Ni}
  \label{fig:g_vrh_ni_fcc_orbital_sup}
\end{subfigure}
\caption{Predicted shear modulus variation ($\text{G'}_{\text{VRH}}$) for a single-atom substitutionally defected Mo and Ni supercell with respect to the undefected one, along the 3d, 4d and 5d series of the periodic table. The black dashed line highlights the pure host matrix case.}
\label{fig:g_vrh_mo_ni_bcc_orbital_sup}
\end{figure}

\begin{figure}[H]
\centering
\begin{subfigure}{.5\textwidth}
  \centering
  \includegraphics[width=.99\linewidth]{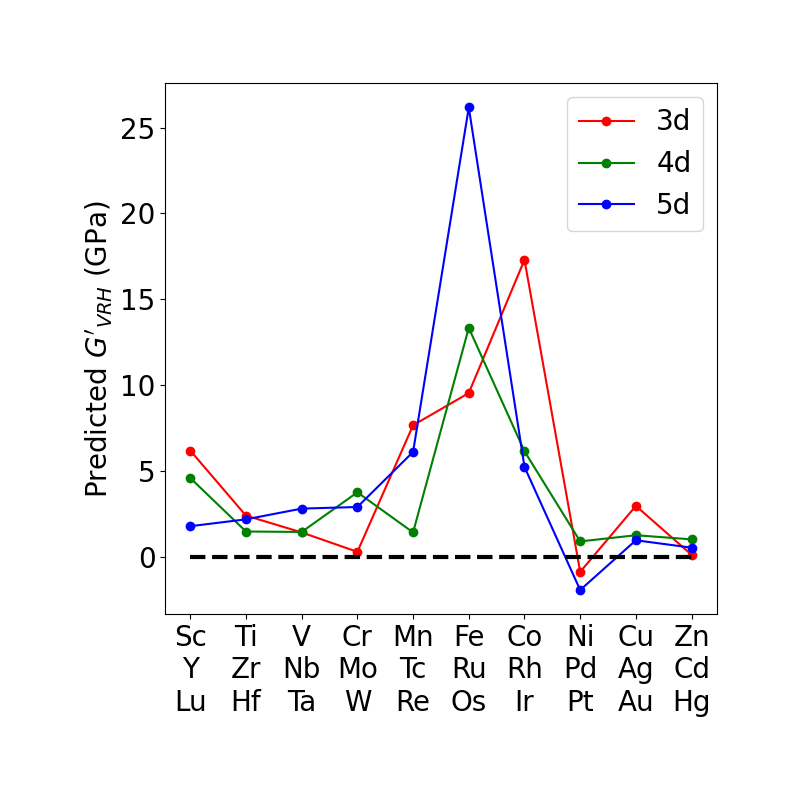}
  \caption{Al}
  \label{fig:g_vrh_al_fcc_orbital_sup}
\end{subfigure}%
\begin{subfigure}{.5\textwidth}
  \centering
  \includegraphics[width=.99\linewidth]{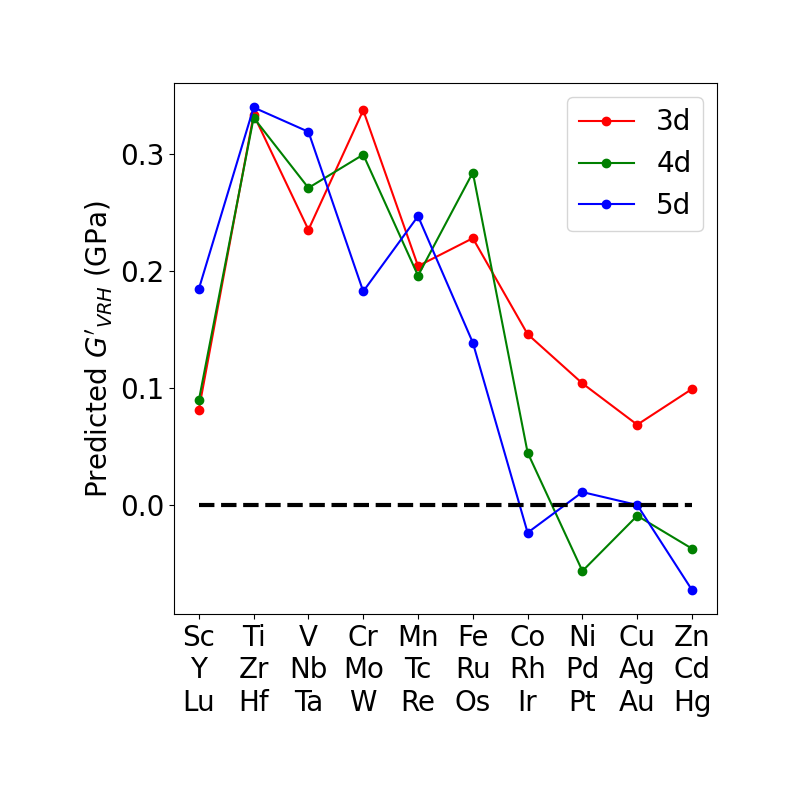}
  \caption{Au}
  \label{fig:g_vrh_au_fcc_orbital_sup}
\end{subfigure}
\caption{Predicted shear modulus variation ($\text{G'}_{\text{VRH}}$) for a single-atom substitutionally defected Al and Au supercell with respect to the undefected one, along the 3d, 4d and 5d series of the periodic table. The black dashed line highlights the pure host matrix case.}
\label{fig:g_vrh_al_au_fcc_orbital_sup}
\end{figure}

\begin{figure}[H]
\centering
\begin{subfigure}{.5\textwidth}
  \centering
  \includegraphics[width=.99\linewidth]{eform_mo_bcc_orbital.png}
  \caption{Mo}
  \label{fig:eform_mo_bcc_orbital_sup}
\end{subfigure}%
\begin{subfigure}{.5\textwidth}
  \centering
  \includegraphics[width=.99\linewidth]{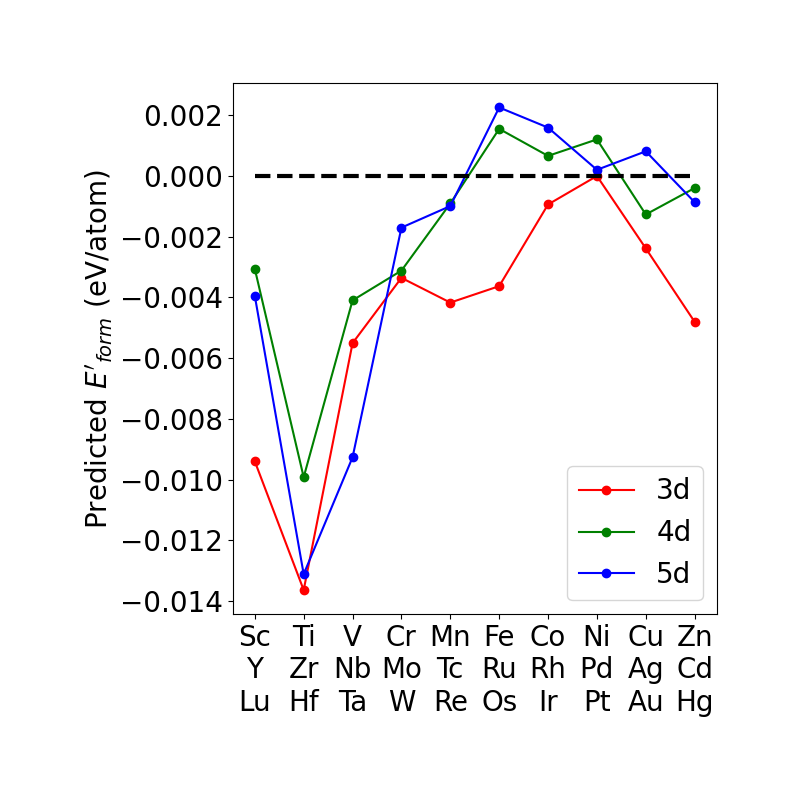}
  \caption{Ni}
  \label{fig:eform_ni_fcc_orbital_sup}
\end{subfigure}
\caption{Predicted formation energy variation ($\text{E'}_{\text{form}}$) for a single-atom substitutionally defected Mo and Ni supercell with respect to the undefected one, along the 3d, 4d and 5d series of the periodic table. The black dashed line highlights the pure host matrix case.}
\label{fig:eform_mo_ni_fcc_orbital}
\end{figure}

\begin{figure}[H]
\centering
\begin{subfigure}{.5\textwidth}
  \centering
  \includegraphics[width=.99\linewidth]{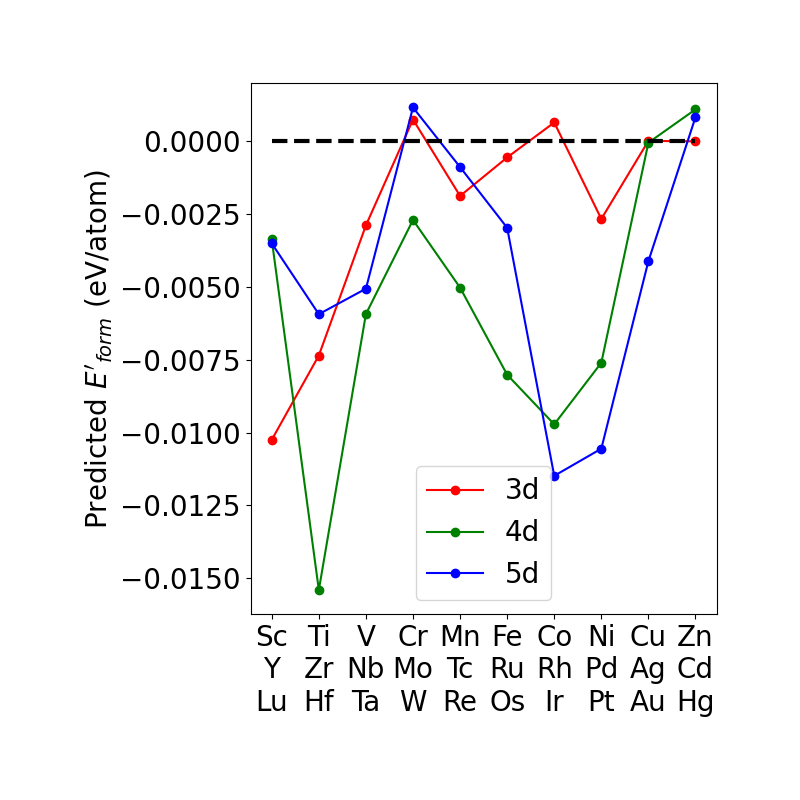}
  \caption{Al}
  \label{fig:eform_al_fcc_orbital_sup}
\end{subfigure}%
\begin{subfigure}{.5\textwidth}
  \centering
  \includegraphics[width=.99\linewidth]{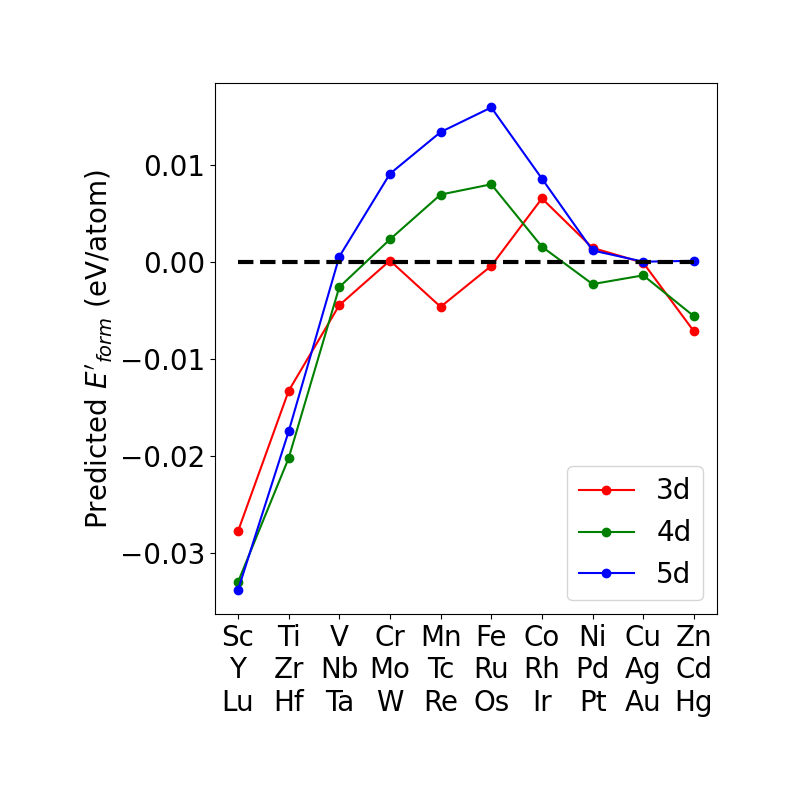}
  \caption{Au}
  \label{fig:eform_au_fcc_orbital_sup}
\end{subfigure}
\caption{Predicted formation energy variation ($\text{E'}_{\text{form}}$) for a single-atom substitutionally defected Al and Au supercell with respect to the undefected one, along the 3d, 4d and 5d series of the periodic table. The black dashed line highlights the pure host matrix case.}
\label{fig:eform_al_au_fcc_orbital_sup}
\end{figure}

\end{appendices}

\end{document}